\documentclass[twocolumn, trackchanges]{aastex701}

\usepackage{graphicx}
\usepackage{amsmath} 
\usepackage{subcaption}
\usepackage{placeins}
\usepackage{longtable}
\usepackage{pgffor}

\begin{document}

\title{Detection Forecasting for Blazars at MeV Energies with Fermi-LAT Swift-BAT Cross Match}

\author[0009-0006-0666-7940]{Timothy Wessling-Resnick}
\affiliation{Department of Physics, Northeastern University, Boston, MA 02115, USA}
\email{wessling-resnick.t@northeastern.edu}

\author{Naomi Tsuji}
\affiliation{
Institute for Cosmic Ray Research, The University of Tokyo, Kashiwanoha 5-1-5, Kashiwa, Chiba 277-8582, Japan
}
\email{}

\author{Tsuguo Aramaki}
\affiliation{Department of Physics, Northeastern University, Boston, MA 02115, USA}
\email{}

\author{Paolo Coppi}
\affiliation{Department of Astronomy, Yale University, PO Box 208101, New Haven, CT 06520-8101, USA}
\email{}

\author{Manel Errando}
\affiliation{Department of Physics, Washington University in St. Louis, St. Louis, MO 63130, USA}
\email{}

\author{Yoshiyuki Inoue}
\affiliation{College of Systems Engineering and Science, Shibaura Institute of Technology, 307 Fukasaku, Minuma, Saitama 337-8570, Japan}
\email{}

\author{Reshmi Mukherjee}
\affiliation{Department of Physics and Astronomy, Barnard College, Columbia University, NY 10027, USA}
\email{}

\author{Svanik Tandon}
\affiliation{Physics Department, Columbia University, New York, NY 10027, USA}
\email{}

\author{Jiancheng Zeng}
\affiliation{Department of Physics, Northeastern University, Boston, MA 02115, USA}
\email{}

\begin{abstract}

Over the past two decades, the lack of an MeV $\gamma$-ray all-sky instrument has left an observational gap in blazar broadband emission. Recently, a new suite of MeV $\gamma$-ray instruments with sensitivity in the $10^{-11}$ to $10^{-12}$ $\mathrm{erg/cm^{2}/s}$ range has been proposed, with the potential to far surpass the capabilities of previous all-sky MeV instruments. In this work, we forecast the capabilities of future MeV $\gamma$-ray instruments including COSI, GRAMS, newASTROGAM, and AMEGO-X to observe blazars. We pair total source count estimates of blazar luminosity functions with SED-modeled flux estimates of 76 blazars selected from a sample of 115 blazars cross-matched between the latest Fermi-LAT and Swift-BAT catalogs. We forecast that an instrument with a sensitivity limit of $10^{-11}$ $\mathrm{erg/cm^{2}/s}$ could detect 61 of the 76 SED-modeled blazars, and expect a total population of a few tens to a few hundreds of Flat-Spectrum Radio Quasars (FSRQs), as well as a few hundred BL Lacs (BLLs). An instrument with a sensitivity limit of $10^{-12}$ $\mathrm{erg/cm^{2}/s}$ could detect all 76 of the SED-modeled blazars, and we expect a total population of a few hundred to one thousand FSRQs and roughly one thousand BLLs. As MeV blazars have been observed to be extremely luminous and MeV $\gamma$-rays are minimally attenuated by the extragalactic background light, we use luminosity functions to estimate that blazars could be detected out to $z\approx6$. Lastly, we demonstrate detectability for blazars that have been proposed as neutrino-emitting candidates, as the MeV regime could provide a window into resolving lepto-hadronic degeneracies. 

\end{abstract}

\section{Introduction}

Blazars are a subclass of radio-loud active galactic nuclei (AGN) with relativistic jets oriented close to the observer's line of sight \citep{Urry_1995}. This orientation results in their observed broadband spectral energy distributions (SEDs) to be dominated by jet emission. Blazars are set apart from other types of AGN by their extreme $\gamma$-ray luminosities. In the $\gamma$-ray sky, blazars are the dominant source type, with the latest Fermi-LAT source catalog containing nearly 4000 blazars, composing over 50$\%$ of point sources and the vast majority of its extragalactic sources \citep{Ballet2023}. Blazars have been widely observed by hard X-ray instruments including INTEGRAL \citep{Integral_Winkler2003}, Swift-BAT \citep{Barthelmy:2005ax}, and NuSTAR \citep{Harrison:2013wda}, GeV $\gamma$-ray instruments including EGRET \citep{Thompson1993} and Fermi-LAT \citep{Atwood:2009ez}, and TeV $\gamma$-ray instruments including VERITAS \citep{Holder2006}. However, the $\gamma$-ray emission of most blazars still lacks full broadband coverage, due to the absence of a high-sensitivity, all-sky MeV $\gamma$-ray instrument. To date, only a handful of instruments have detected sources within the MeV band (0.1--100 MeV). The foremost of these instruments was the Imaging Compton Telescope (COMPTEL), which operated through the 1990s \citep{Schoenfelder2000COMPTEL}. With a sensitivity of roughly $10^{-10}$ erg~cm$^{-2}$~s$^{-1}$, COMPTEL detected nine blazars in the 0.75--30 MeV range. Since the end of COMPTEL's operations, there has yet to be a dedicated MeV all-sky instrument, leading to what many have dubbed the ``MeV gap'' in broadband studies of astrophysical sources. Some hard X-ray instruments, such as Swift-BAT, have detected blazars at energies up to $\sim0.2$ MeV, and Fermi-LAT has detected blazars with greater than $\sim$5$\sigma$ significance in the 20--200 MeV band \citep{FermiLAT2026_2FLE}. However, a future MeV $\gamma$-ray instrument is required to bridge the MeV gap and achieve complete broadband $\gamma$-ray coverage for blazars. 

\vspace{0.5 em}

\noindent Blazars are traditionally classified into flat-spectrum radio quasars (FSRQs) and BL Lacs (BLLs) based on the equivalent width of their optical lines \citep{Stickel1991}. Blazars lacking optical line information are classified as Blazars of Classification Unknown, or BCUs in the Fermi-LAT catalog \citep{Abdollahi_2020}. The general shape of blazar broadband SEDs is characterized by two non-thermal humps. The lower-energy hump, spanning from radio to X-ray energies, originates from synchrotron radiation. The high-energy hump, ranging from hard X-ray to $\gamma$-ray energies, is primarily attributed to the inverse-Compton (IC) scattering of low energy photons by relativistic charged particles. The peak frequency of the synchrotron component has been used as an alternative classification scheme, sorting blazars into low-, intermediate-, and high-synchrotron peaked (LSP, ISP, HSP) classes \citep{Abdo_2010}. The vast majority of FSRQs are LSP-type blazars, while BLLs have been identified with each synchrotron classification \citep{Ballet2023}.

\vspace{0.5em}

\noindent In the $\gamma$-ray band, the spectral peak of the high-energy component has been shown to occur in the MeV band for the most luminous class of blazars, predominantly FSRQs. These types of blazars have been dubbed ``MeV blazars'' \citep{Sikora:2002zb, Sbarrato_2013, Marcotulli_2020}. MeV blazars have thus far been observed to be extremely luminous, and MeV $\gamma$-ray photons are much less attenuated than their GeV counterparts \citep{Venters_2009, Ackermann2012EBL, Dwek2013}. As a result, future MeV $\gamma$-ray instruments may be able to search for high-redshift blazars, and serve as probes of the history of relativistic jets through cosmic time. For blazars with lower $\gamma$-ray luminosities such as BLLs, the spectral peak has been shown to occur at GeV and TeV energies, with the valley between the synchrotron and inverse-Compton emission components situated in the MeV $\gamma$-ray regime. The reddening of the energy and increase in luminosity luminosity of the high-energy spectral peak has been attributed to differences in accretion-disk luminosity \citep{Ghisellini2009}. More powerful accretion disks in FSRQs enhance external-Compton contributions to the seed photon field off which relativistic particles scatter. This leads to more frequent scattering events, but lower-energy scattered photons. The blazar sequence \citep{Ghisellini1998_unifying, Ghisellini2017_FermiSequence} proposes an anti-correlation between bolometric $\gamma$-ray luminosity and the high-energy spectral peak. This is consistent with FSRQs and BLLs representing a unified class of objects, where BLLs are accretion-starved FSRQs. However, there is some disagreement on whether this trend is due to a selection effect, so an MeV instrument is necessary to determine the precise energy of the high-energy spectral peak \citep{Ajello_2012, Marcotulli:2022BASS33}. 

\vspace{0.5em}

\noindent Different blazar SED models invoke distinct mechanisms for $\gamma$-ray production, and thus observation by future MeV $\gamma$-ray instruments is required to constrain contributing jet processes. In particular, blazars are thought to be candidate sources for the production of astrophysical neutrinos, with the IceCube-170922A neutrino event associated at a significance of $\sim3\sigma$ with the $\gamma$-ray blazar TXS 0506+056 \citep{IceCube:2018dnn} as the strongest evidence thus far. Neutrinos could be produced by hadronic or photo-hadronic interactions in blazar jets that also produce $\gamma$-rays through neutral pion decay \citep{Cerruti2018, Boettcher:2019mma, Sbarrato:2025mev}. Neutrinos may also be generated by hadronic sources in the disk corona of AGN, with evidence for this supported by the association of neutrinos with the Seyfert galaxy NGC 1068 \citep{IceCube2022NGC1068}. In both scenarios, $\gamma$-rays produced in the same hadronic processes as neutrinos are expected to cascade down to MeV or hard X-ray energies \citep{Inoue2020}. Future MeV $\gamma$-ray instruments could observe the imprint of these cascade photons, thus providing a signature for hadronic processes in blazar jets. Given the nature and origin of astrophysical neutrinos is still highly debated, an MeV all-sky instrument will be uniquely positioned to offer insight to these processes. While FSRQs are thought to be more efficient at producing neutrinos, the signature of the hadronic component may be buried under the leptonic emission \citep{Atoyan2003}. Conversely, BLLs are thought to be less efficient at producing neutrinos, but the valley in which the MeV region lies may be ideal for the hadronic component to be observed. In addition, the polarization of leptonic and hadronic models are expected to be different, with leptonic emission expected to be less polarized than hadronic emission. Proposed next-generation MeV $\gamma$-ray instruments may have polarization capabilities, and thus if polarization in MeV $\gamma$-ray emission is measured and coincident with neutrino alerts, it could be strong evidence for hadronic processes \citep{Zhang2013XrayGammaPol}. 

\vspace{0.5 em}

\noindent The upcoming Compton Spectrometer and Imager (COSI) \citep{Tomsick2019_COSI} will be the first instrument to survey the sky with sensitivity beyond that of COMPTEL. COSI will survey the sky in the 0.2--5 MeV range, opening the door on a new era of MeV $\gamma$-ray science. Following COSI, a new suite of next-generation instruments has been proposed to survey the entire MeV regime, and improve sensitivity by at least an order of magnitude relative to COMPTEL. These instruments include the All-sky Medium-Energy Gamma-ray Observatory (AMEGO) \citep{McEnery:2019tcm, Perkins:2022amegox}, the ASTROnomical Gamma-ray Analysis Mission (ASTROGAM) \citep{DeAngelis2018_eASTROGAM, Rando_2019, Berge:2025newastrogam}, and the Gamma-Ray and AntiMatter Survey (GRAMS) \citep{Aramaki2020_GRAMS}. Other instruments including SMILE \citep{SMILE3_Takada}, GECCO \citep{Moiseev2023}, and MeVCube \citep{Lucchetta2022} are also designed to observe the MeV $\gamma$-ray regime. These instruments will largely probe the 0.1--10 MeV regime, and thus we will primarily focus on instruments that will be sensitive across a more complete span of the MeV $\gamma$-ray regime.

\vspace{0.5em}

\noindent In this work, we present a novel forecast for the capabilities of next-generation MeV $\gamma$-ray instruments to observe blazars. Previous work has used luminosity-function models to estimate the cumulative blazar source counts in the MeV band \citep{Inoue2015MeVAGN, Marcotulli:2022BASS33}. We complement this upper limit of expected source counts with a robust lower limit from SED modeling of a sample of blazars cross-matched between the latest data releases from Fermi-LAT and Swift-BAT. This paper is organized into the following sections. In Section \ref{sec:multiwavelength_data}, we will present the methodology and results of cross-matching. In Section \ref{sec:cm_blazars}, we will present cross-matched blazar sample properties and phenomenological modeling of cross-matched blazar SEDs. In Section \ref{sec:detection_forecasting}, we will present the estimated counts that can be expected from cross-matched SED modeling and luminosity-function-based cumulative source count estimates for different orders of magnitude sensitivity. We will then present our forecast for the sensitivities of future MeV $\gamma$-ray instruments, including COSI, newASTROGAM, AMEGO-X, and GRAMS. In Section \ref{sec:high_z_forecasting}, we will discuss the potential for high-redshift detections and the redshift limits at which future MeV $\gamma$-ray missions may be capable of detecting blazars. Lastly, in Section \ref{sec:discussion}, we will discuss the potential for neutrino multimessenger studies by MeV $\gamma$-ray instruments. In this work we adopt a flat cosmology with $\Omega_m = 0.3$, $\Omega_{\Lambda}=0.7$, and $H_0 = 70$ $\mathrm{km/s/Mpc}$.   

\section{Blazar Cross Matching}
\label{sec:multiwavelength_data}

We apply the methodology described in \citet{Tsuji_2021} to cross-match the latest catalog data releases from the Swift-BAT hard X-ray survey and Fermi-LAT GeV $\gamma$-ray survey to identify a population of blazars with both hard X-ray and GeV $\gamma$-ray coverage.

\subsection{Swift-BAT 157-Month Catalog}

The most recent data release of the Swift-BAT survey is the 157-month Swift-BAT catalog \citep{Lien_2025} (henceforth BAT; other data releases of the Swift-BAT survey will be explicitly specified).  This release contains 1891 point sources, a roughly 16$\%$ increase from the 1623 sources in the 105-month Swift-BAT catalog \citep{Oh_2018} used in \citet{Tsuji_2021}. In this release of the Swift-BAT catalog, many of the BAT blazar classifications follow the scheme of the RomaBZCAT \citep{Massaro2015_RomaBZCAT5}, where BZQs are FSRQs, BZBs are BLLs, and BZGs are blazars where emission from the host galaxy can be observed in the source SED. Beamed AGN serves as a general classifier for blazars in BAT. Other classifications used in this release include FSRQ, BZQ/Lense, and Blazar. Of the total point source population in the BAT catalog, 184 sources are blazars. In the BAT 105-month survey catalog, the classification of Beamed AGN was used to encompass all types of blazars, comprising 158 of the total sources. Thus, the number of blazars in the BAT 157-month catalog marks a roughly 16$\%$ increase in comparison to the BAT 105-month catalog, similar to the increase in total sources.  

\subsection{Fermi-LAT 4FGL-DR4 Catalog}

The most recent data release from the Fermi-LAT survey is 4FGL-DR4, the fourth data release of the fourth Fermi-LAT source catalog \citep{Ballet2023, Abdollahi_2020} (henceforth LAT; other data releases of the Fermi-LAT catalog will be explicitly specified). This release contains 7195 point sources. LAT also contains extended sources, which were matched in \citet{Tsuji_2021} for the second data release of the fourth Fermi-LAT source catalog (4FGL-DR2) \citep{Ballet2020_4FGL_DR2}. However, none of the extended sources listed in either LAT catalog data release are blazars, and thus we do not discuss the matching of extended sources in this work. In LAT, blazars are classified as FSRQ/fsrq, BLL/bll, and BCU/bcu, denoting identified and associated sources \citep{Abdollahi_2020}. Of the point sources contained in 4FGL-DR4, 3933 are classified as blazars. 4FGL-DR2 contained 5788 sources, of which 3436 were classified as blazars, thus marking a 24$\%$ increase in the number of point sources and a 14$\%$ increase in the number of blazars.

\subsection{4LAC-DR3 Catalog}

We also utilize additional information presented in 4LAC-DR3, the third data release of the fourth Fermi AGN catalog \citep{Ajello_2022} (henceforth LAC; other data releases of LAC will be explicitly specified). LAC is a supplemental catalog to the third data release of the Fermi-LAT source catalog \citep{Abdollahi_2022}. This catalog contains 3743 blazar sources, of which 3737 are also listed in LAT. For the sources listed in both catalogs, we utilize the SED classification and the listed redshifts.

\subsection{Matching}

In this work, we utilize the methodology of \citet{Tsuji_2021} to cross-match the latest LAT and BAT catalogs. By doing so, we identify a sample of matched blazars for which we can forecast observability. We note that while we complete a full cross-match of the latest BAT and LAT catalogs, this work discusses only matched blazars. The complete updated cross-match inclusive of all source types will be presented in a forthcoming work. \citet{Tsuji_2021} conducted cross-matching between the Swift-BAT 105-month source catalog of the second data release of the fourth Fermi-LAT source catalog. To match sources \citet{Tsuji_2021} first conducted spatial matching, applying a limit of $0.08^\circ$ to the separation of source coordinates. Spatial matching was then followed up by identification matching, which identified sources with the same association names in both catalogs. The quality of each match was evaluated using a simple flagging system, with firm matches flagged as $M$ to be those matches identified to have the same source type and name (see Section 4.1 of \citet{Tsuji_2021} for a full review of match flags).

\vspace{0.5 em}

\noindent To ensure certainty of blazar classification for sources in our sample, we select only matches that are classified as blazars in both LAT and BAT and share the same association name (blazars flagged as firmly matched). We first conduct spatial matching of the BAT and LAT catalogs, applying the same spatial separation limit as \citet{Tsuji_2021} of $r_{sep} < 0.08^\circ$. Through spatial matching, we find a total of 96 matches where both the BAT and LAT counterparts are blazars. Through identification matching, we find an additional 19 matches where both BAT and LAT counterparts are blazars, thus bringing the sample total to 115 firmly matched blazars. \citet{Tsuji_2021} identified 96 firmly matched blazars out of 156 total matched sources. Thus, we find a roughly 20$\%$ increase in the number of matched blazars in this work. This result is expected given the respective increases in BAT and LAT blazars in the latest data releases. The complete list of matched sources is shown in Table \ref{tab:MeV_Blazars} in Appendix \ref{sec:MeV_Blazar_catalog}. We report the LAT and BAT sources by their association names as listed in the LAT \texttt{ASSOC1} catalog column and the BAT \texttt{COUNTERPART$\_$NAME} column.  

\vspace{0.5 em}

\noindent Through spatial matching, we also find 15 matches where only one of the counterparts was identified as a blazar. Of these, 12 LAT blazars had non-blazar BAT counterparts, consisting of 7 Seyfert galaxies, 2 unknown sources, 1 Other AGN, 1 confused source, and 1 cataclysmic variable star (CV), while 3 BAT blazars had non-blazar LAT counterparts, consisting of 2 compact steep-spectrum radio sources (css) and 1 narrow-line Seyfert 1 galaxy. All identification-matched blazars were firmly matched.

\vspace{0.5 em}

\noindent In the latest data release of BAT, the classification scheme for AGN was revised relative to the 105-month survey, resulting in the reclassification of many AGN. As a result, two matches, PMN J1310-5552 (LAT) with IGR J13109-5552 (BAT) and PKS 2331-240 (both BAT and LAT), which were not considered firmly matched blazars in \citet{Tsuji_2021}, are now considered firmly matched. The BAT counterpart IGR J13109-5552 was reclassified as BZQ from Sy1. For PKS 2331-240, the BAT counterpart was reclassified as a BZQ from a beamed AGN, and the LAT counterpart was reclassified as a bcu from an agn. Likewise, one match, PKS 0524-460 (both BAT and LAT), had its BAT counterpart reclassified as Sy1 from beamed AGN, and is thus no longer firmly matched. It is now flagged as $D$ (same name, different type; see \citet{Tsuji_2021}). The remaining 18 firmly matched blazars were not previously identified in \citet{Tsuji_2021}. In the remainder of this work, we primarily refer to matched sources by their LAT association name and LAT classification, as the LAT classification for blazars follows the typical optical classification methodology \citep{Stickel1991}.

\begin{deluxetable*}{lcccc}[t]
\tablecaption{Classification Distributions \label{tab:class_dist}}
\tablehead{
\colhead{Class} & \colhead{$N_{matched}$\tablenotemark{a}} & \colhead{$\%_{matched}$\tablenotemark{b}} & \colhead{$N_{catalog}$\tablenotemark{a}} & \colhead{$\%_{catalog}$\tablenotemark{b}}
}
\startdata
\textbf{BAT Classification} & & & & \\
BZQ & 44 & 38$\%$ & 74 & 40$\%$\\
Beamed AGN & 32 & 28$\%$ & 57 & 31$\%$\\
BZB & 22 & 19$\%$ & 22 & 12$\%$\\
FSRQ & 10 & 9$\%$ & 20 & 11$\%$\\
BZG & 6 & 5$\%$ & 8 & 4$\%$\\
Blazar & 0 & $0\%$ & 2 & 1$\%$ \\
BZQ/Lense & 1 & 1$\%$ & 1 & 1$\%$\\
\textbf{LAT Classification} & & & & \\
FSRQ/fsrq & 61 & 53$\%$ & 820 & 21$\%$\\
BLL/bll & 43 & 37$\%$ & 1490 & 38$\%$\\
BCU/bcu & 11 & 10$\%$ & 1623 & 41$\%$\\
\textbf{LAC SED Classification} & & & & \\
LSP & 69 & 60$\%$ & 1664 & 42$\%$\\
ISP & 2 & 2$\%$ & 522 & 13$\%$\\
HSP & 38 & 33$\%$ & 584 & 15$\%$\\
None & 6 & 5$\%$ & 1163 & 30$\%$
\enddata
\tablenotetext{a}{$N_{matched/catalog}$ -- Number of sources of each classification in the matched and parent catalogs.}
\tablenotetext{b}{$\%_{matched/catalog}$ -- Percentage of sources of each classification in the matched and parent catalogs.}
\end{deluxetable*}
\label{tab:source_dist}

\subsubsection{Redshift Validation}

Of the 115 firmly matched sources, 113 of their LAT counterparts are listed in LAC. Given that redshift is an important characteristic for blazars, we compare the available redshifts listed in LAC and BAT as a further validation of firmly matched blazars. Of the 115 firmly matched blazars, 100 of their LAT counterparts have redshifts listed in LAC, and 98 of their BAT counterparts have known redshifts. Of these matches, 94 have redshifts available for both their BAT and LAT counterparts, and thus we compare the BAT and LAT redshifts as an additional method for verifying the match. Of the 94 matches where both BAT and LAC redshifts are available, only one match, PMN J1310-5552 (LAT) with IGR J13109-5552 (BAT), was found to have a large disparity in redshift measurements. \citet{Ajello_2022} lists the redshift of PMN J1310-5552 to be 0.104, while \citet{Lien_2025} lists the redshift of IGR J13109-5552 to be 1.559. \citet{Paliya_2024} attributes the discrepancy to confusion in identifying the correct optical counterpart, due to the source's proximity to the Galactic plane. The redshift listed for PMN J1310-5552 in \citet{Ajello_2022} is consistent with that presented by \citet{Masetti2008}, which reported stellar absorption features in the optical spectrum. A later optical campaign conducted by \citet{Koss_2022} found the optical spectrum of PMN J1310-5552 to be AGN-like, reporting the redshift to be 1.56. Since our matching relies solely on the positional separation and classification of the hard X-ray and GeV $\gamma$-ray counterparts, this discrepancy does not affect the validity of the match. Given that the redshift reported in \citet{Lien_2025} matches the redshift derived from the AGN-like spectrum in \citet{Koss_2022}, we adopt the BAT redshift of $z=1.559$ for this match.

\vspace{0.5em}

\noindent In future work, similar validation studies should be conducted when large firmly matched blazar BAT and LAT redshift discrepancies occur. For 10 matches, only one of the counterparts has a listed redshift, and thus we simply adopt the known redshift. For the 11 remaining matches, neither counterpart has a redshift available.

\section{Cross-Matched Blazar Sample}
\label{sec:cm_blazars}
\subsection{Sample Properties}
\label{sec:cm_properties}

\begin{figure}[h]
    \centering
    \includegraphics[width=\linewidth]{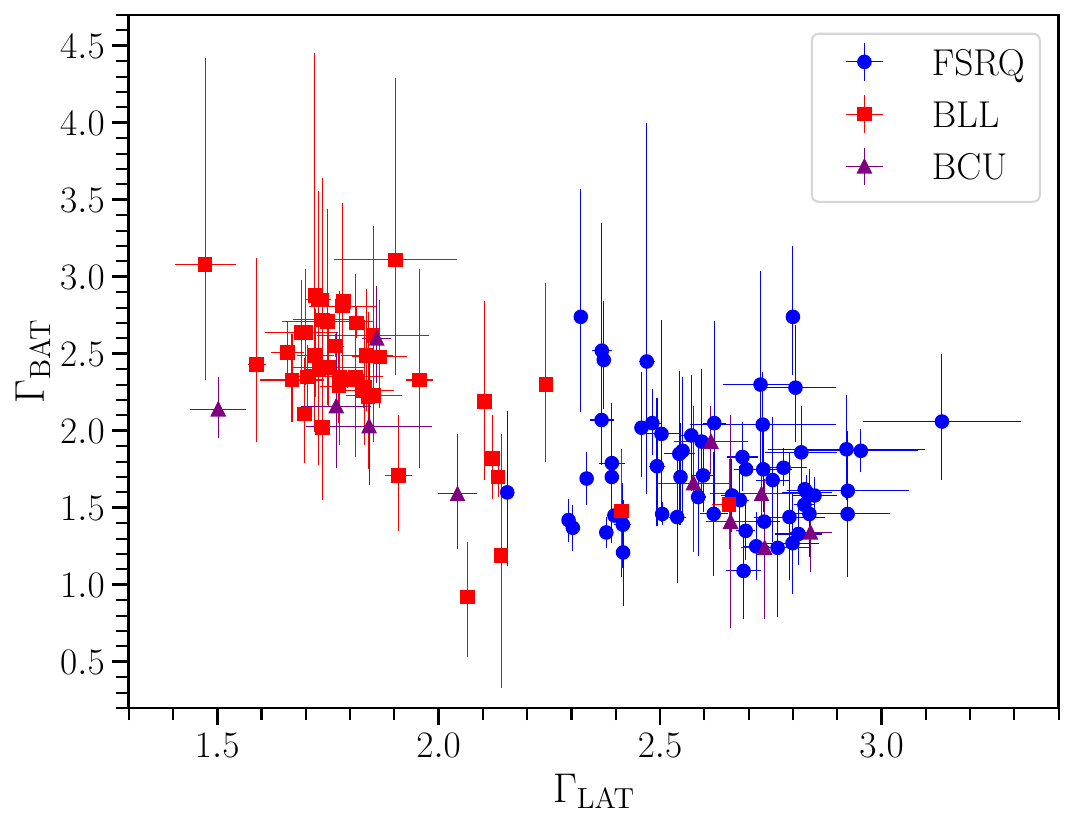}
    \caption{The distributions of $\Gamma_{\mathrm{BAT}}$ (14 keV--195 keV) and $\Gamma_{\mathrm{LAT}}$ (100 MeV--1 TeV) for all firmly matched blazar sources. Note that for this comparison we use the photon index from \cite{Ballet2023} that was derived using a power law.}
    \label{fig:Matched_Spectal_indexes}
\end{figure}

Table \ref{tab:class_dist} shows the classification distributions of the firmly cross-matched blazars in comparison with the distribution of blazar classifications in their parent BAT, LAT, and LAC catalogs. The population of firmly matched blazars, is composed primarily of LSP-type FSRQs and HSP-type BLLs. These two groups of sources encompass 56 and 34 respectively of the 115 firmly matched blazars, or roughly 78$\%$ of the total matched population. The remaining types of matched sources include 7 LSP-type BLLs, 6 LSP-type BCUs, 5 FSRQs with no SED classification, 4 HSP-type BCUs, 2 ISP-type BLLs, and 1 BCU with no SED classification. Figure \ref{fig:Matched_Spectal_indexes} compares the power-law spectral indices from BAT and LAT, demonstrating that these two populations have distinct high-energy spectral shapes. LSP-type FSRQs have soft LAT indices, with an average $\Gamma_{\mathrm{LAT}}$ of 2.395 $\pm$ 0.002 and hard BAT indices with an average of $\Gamma_{\mathrm{BAT}}$ of 1.681 $\pm$ 0.012. Conversely HSP-type BLLs have hard LAT indices, with a $\Gamma_{\mathrm{LAT}}$ of 1.768 $\pm$ 0.003 and soft BAT indices with an average of $\Gamma_{\mathrm{BAT}}$ of 2.611 $\pm$ 0.024. LSP-type BLLs have average spectral indices similar to LSP-type FSRQs, with an average $\Gamma_{\mathrm{LAT}}$ of 2.152 $\pm$ 0.003 and $\Gamma_{\mathrm{BAT}}$ of 1.705 $\pm$ 0.101. The spectral index properties of the different populations in our sample motivate our methodology for phenomenological spectral energy distribution (SED) modeling. We note that the cross-matched sample is intrinsically biased towards bright sources, with BAT fluxes (14-195 keV) exceeding $10^{-12}$ erg~cm$^{-2}$~s$^{-1}$ and LAT fluxes (50 MeV to 1 TeV) exceeding 6$\times10^{-13}$ erg~cm$^{-2}$~s$^{-1}$. The Swift-BAT survey is less sensitive than Fermi-LAT, and thus most sources in BAT are relatively bright. Among these bright sources, the cross-matched sample contains 6 blazars that were detected by COMPTEL \citep{Schonfelder2000}. These blazars include 3C 273, 3C 279, 3C 454.3, CTA 102, PKS B1622-297, and PKS 0528+134.

\subsection{Phenomenological SED Modeling}
\label{sec:phenom_model}

To estimate the fluxes in the MeV regime for our firmly matched blazars, we conduct phenomenological modeling of the multiwavelength time-averaged SED data from BAT and LAT. BAT and LAT spectral data are binned into eight spectral bands: 14--20, 20--24, 24--35, 35--50, 50--75, 75--100, 100--150, and 150--195 keV; and 0.05--0.1, 0.1--0.3, 0.3--1, 1--3, 3--10, 10--30, 30--100, and 100--1000 GeV, respectively. To extrapolate across the MeV gap for matched blazars, we utilize simple phenomenological SED models. We fit all FSRQs with a log parabola (LP) model as an approximation for a one-zone leptonic model frequently used to describe the high-energy emission from FSRQs \citep{Ghisellini2009}. The softer LAT spectral indices and harder BAT spectral indices of FSRQs are indicative of the presence of a spectral peak in the MeV energy regime, and is thus consistent with this choice of model. LSP-type BLLs exhibit similar spectral properties and thus can also be appropriately modeled with the LP model. We fit the LP model shown in Equation \ref{eq:LP} utilizing \texttt{gammapy} \citep{gammapy:2023}. $N_{0}$ is set to be the differential flux of the 14--20 keV BAT bin, and $E_0$ is set to be the bin center of 17 keV, $\alpha$ is the photon index, and $\beta$ is the curvature.

\begin{equation}
    \frac{dN}{dE} = N_{0} \left( \frac{E}{E_{0}} \right)^{-\alpha-\beta  \log \left( \frac{E}{E_{0}} \right)}
    \label{eq:LP}
\end{equation}

\noindent For HSP-type BLLs, their softer hard X-ray indices and harder GeV $\gamma$-ray indices are consistent with the falling edge of the synchrotron component transitioning into the rising edge of the inverse-Compton component. A simple phenomenological extrapolation across the MeV region of the SED for this class of blazars would be to use a smoothed-broken-power-law (SBPL). However, the spectral shape and flux in this region is highly dependent on the hardness and curvature of the falling edge of the synchrotron component and the rising edge of the inverse-Compton component. Depending on the nature of the spectral shape, the flux of this class of sources in the MeV regime could vary by multiple orders of magnitude. Thus, we show the SBPL as an optimistic model for the spectral shape of valley sources, but do not include their modeled fluxes in our forecast, as we may overestimate the number of expected sources. For this model, $E_{0}$ and $N_{0}$ are the same as defined for the LP model, and $\Gamma_{1}$ is the photon index when $E>E_{break}$, and $\Gamma_{2}$ is the photon index when $E<E_{break}$. $\beta$ is the curvature around the break energy, which we fix to be $\beta$ = 1. 

\begin{equation}
\frac{dN}{dE} = N_{0} 
\left( \frac{E}{E_{0}} \right)^{-\Gamma_{1}} 
\left[ 1 + \left( \frac{E}{E_{\rm break}} \right)^{\frac{\Gamma_{2} - \Gamma_{1}}{\beta}} \right]^{-\beta}
\label{eq:SBPL}
\end{equation}

\begin{figure*}[t]
\centering

\includegraphics[width=0.99\textwidth]{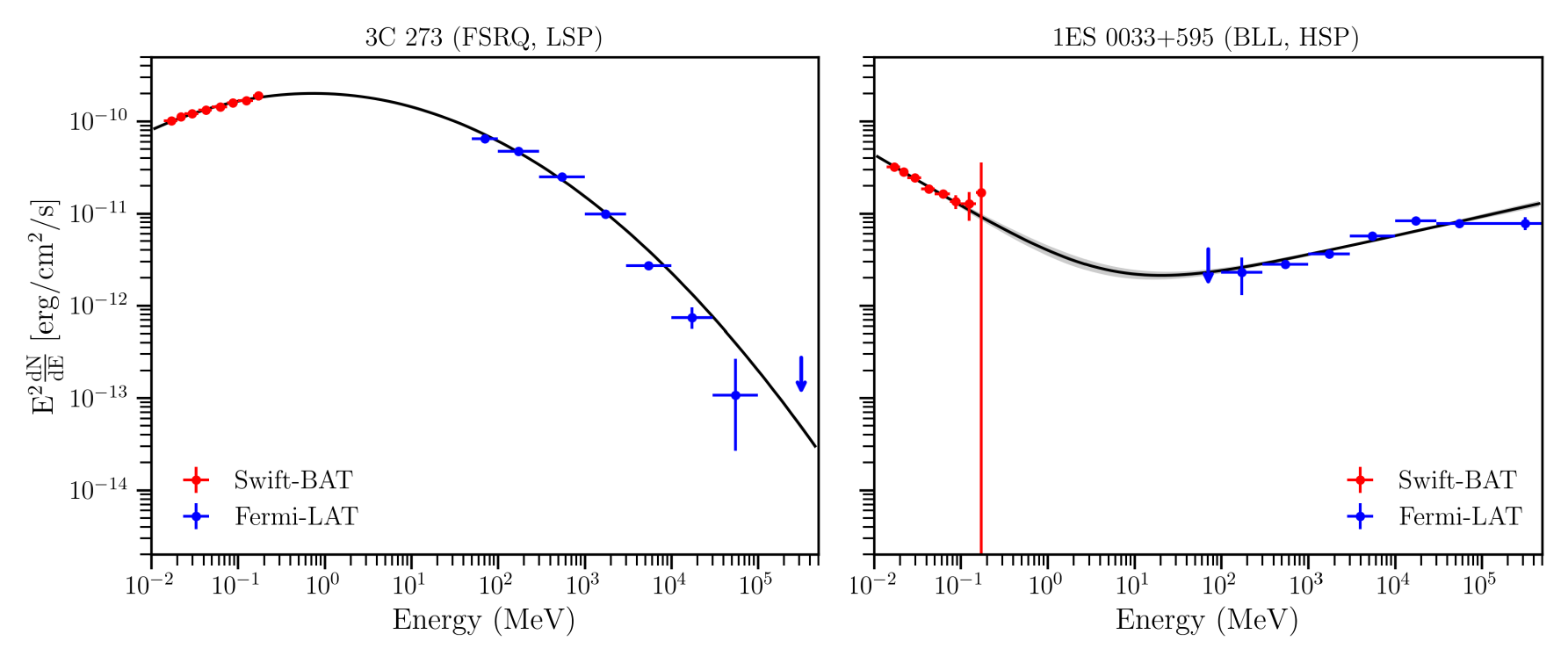}

\caption{The left panel shows the SED of the LSP FSRQ 3C 273, fit with the log parabola model ($N_0$ = 9.9 $\times10^{-11}$ erg~cm$^{-2}$~s$^{-1}$, $\alpha$ = 1.6, $\beta$ = 0.05), and the right panel shows the SED of 1ES 0033+595 fit with the smoothed-broken-power-law-model ($N_0$ = 3.2 $\times10^{-11}$ erg~cm$^{-2}$~s$^{-1}$, $\Gamma_1$ = 2.6, $\Gamma_2$ = 1.8, $E_{break}$ = 5.7 MeV).}
\label{fig:example_SEDs}
\end{figure*}

Similarly to LSP-type BLLs and FSRQs, the SED of BCUs with softer LAT spectral indices and harder BAT spectral indices is well described by the LP model. One HSP-type BCU, PMN J1310-5552, has a hard LAT spectral index of 2.73 and a soft BAT index of 1.59. This source is notably classified as a BZQ (FSRQ) in BAT, and thus may belong to a rare class of HSP-type FSRQs. Given its spectral shape, the SED of this source is well described by the LP. The ISP type BLL S5 0716+71 also has a soft BAT index and hard LAT index, and thus we select the LP model. 

\vspace{0.5em}

\noindent To evaluate the quality of our SED models, we utilize a reduced $\chi^{2}$ test (delta degree of freedom = 2 for LP, 3 for SBPL). For sources where we select the LP model, the average $\chi^2$ test score is 3.40. As a robustness test to validate our model selection, we evaluate the fit quality if we were to swap the model selection. If we fit all sources where we selected the LP model with the SBPL, we find the average $\chi^2$ score to be 7.27. In total, the LP model was selected for 76 sources, of which 61 are FSRQs, 8 are BLLs, and 7 are BCUs. For these 76 of 115 cross-matched source sample, we will utilize the flux estimates to make forecasts. For valley sources, we do not include the flux estimates made with the SBPL model. 

\vspace{0.5em}

\noindent Figure \ref{fig:example_SEDs} shows the SED models for two sources, 3C 273 (left panel of Figure \ref{fig:example_SEDs}) with the LP model, and 1ES 0033+595 (right panel of Figure \ref{fig:example_SEDs}) with the SBPL model. We note that while spectral data for COMPTEL blazars in the MeV regime is available, many these data points are reported only as upper limits, or during flaring states. Due to this, the use of COMPTEL data does not provide robust validation for the spectral shapes of our sources in the MeV energy regime. 

\section{Blazar Detection Forecasting}
\label{sec:detection_forecasting}

Utilizing flux estimates made with SED modeling of cross-matched blazars, we can set a robust, data-driven lower limit on the detection capabilities of future MeV $\gamma$-ray instruments to observe blazars. We combine this lower limit with cumulative source-count predictions based on estimates made with luminosity functions. This is similar to work done in previous studies such as \citet{Inoue2015MeVAGN} and \citet{Marcotulli:2022BASS33}, which set a model-driven upper limit on the total blazar population in the MeV band. In this work, we make forecasts at nine selected energies across the MeV band: 0.2, 0.5, 1.0, 2.0, 5.0, 10.0, 20.0, 50.0, and 100 MeV. In Sections \ref{sec:detection_counts} and \ref{sec:N_S_counts}, we present the number of detectable cross-matched sources and expected source counts based on luminosity functions at different magnitudes of sensitivity. In Section \ref{sec:forecast}, we present the counts expected based on the sensitivity curves of future MeV $\gamma$-ray instruments. We define a source as detectable if its flux exceeds the sensitivity required for a 3$\sigma$ significance detection at a given energy for both arbitrarily chosen sensitivity values or for sensitivities derived from instrument-specific sensitivity curves. We select $3\sigma$ as this is the typical detection confidence level at which instrument sensitivity curves are defined.

\subsection{Cross-Matched Blazar Detection Forecasting}
\label{sec:cm_forecasting}

\subsubsection{Extragalactic Skymaps}
\label{sec:skymap}

\begin{figure*}[t]
    \centering
    \includegraphics[width=0.99\textwidth]{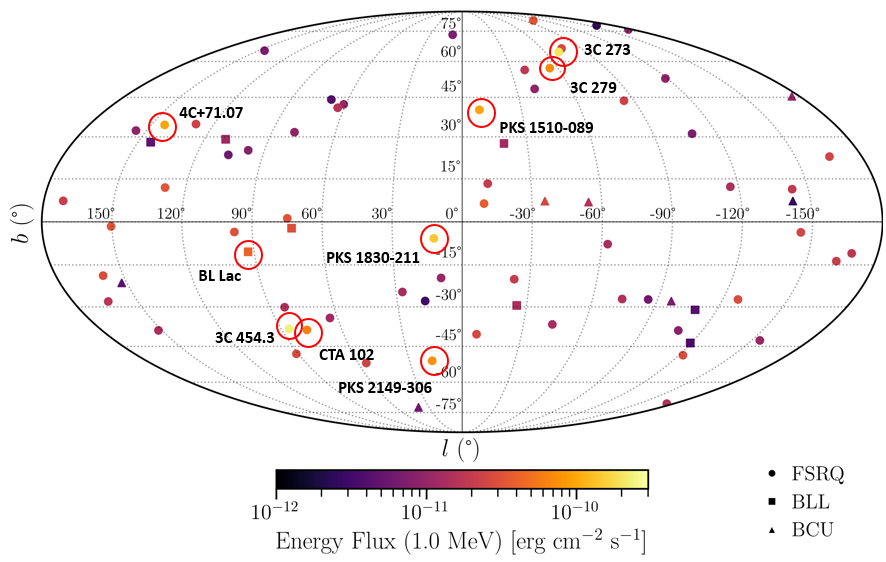}
    \caption{Skymap showing all firmly cross-matched blazar point sources with their fluxes at 1 MeV visualized on a colorbar scale. FSRQs are marked with a circle, BLLs are marked with a square, and BCUs are marked with a triangle. We highlight the 9 brightest sources in red circles with their Fermi counterpart names.}
    \label{fig:MeV_skymap}
\end{figure*}

Figure \ref{fig:MeV_skymap} shows a skymap including the cross-matched blazars for which make flux estimates at 1 MeV. We highlight the nine brightest sources at 1 MeV: 3C 454.3, 3C 273, PKS 1830-211, 4C +71.07, PKS 1510-089, PKS 2149-306, 3C 279, CTA 102, and BL Lac. These sources are all known to be bright sources, exceeding fluxes of $10^{-11}$ $\mathrm{erg/cm^{2}/s}$ at 1 MeV, and remain largely bright across all energies. Four of these sources, including 3C 454.3, 3C 273, 3C 279, and CTA 102, were notably also detected by COMPTEL \citep{Schonfelder2000}. Eight of the nine brightest sources are FSRQs, with the only BLL among them being BL Lacertae. These sources will serve as strong candidates of extragalactic observation targets for early MeV science missions. 3C 454.3 is particularly notable as it is the brightest source at seven out of nine energies and the second brightest source at 0.2 and 0.5 MeV. Other sources, such as 3C 273, are much brighter at lower energies, and fainter at higher energies. 

\vspace{0.5em}

\noindent It should be noted that future MeV instruments will have poor spatial resolution, between 3--5$^\circ$ \citep{Aramaki2020_GRAMS, Perkins:2022amegox, Tomsick2023COSI, Berge:2025newastrogam}. Thus, it will be critical to demonstrate the ability to resolve sources in close proximity. 3C 454.3 and CTA 102 are two bright sources in close proximity on the sky, and thus may be ideal targets for determining resolvability. 3C 273 and 3C 279 are similarly close, and 3C 273 is extremely close to the fainter FSRQ PKS 1329-049. This limitation will be further discussed in Section \ref{sec:cross_match_discussion}.

\subsubsection{Cross-Matched Detection Counts}
\label{sec:detection_counts}

The sensitivity limit required to detect a single source at any of our selected energies is 3.9 $\times$ $10^{-10}$ erg~cm$^{-2}$~s$^{-1}$, and the sensitivity required to detect all matched blazars at all selected energies is 4.4 $\times$ $10^{-13}$ erg~cm$^{-2}$~s$^{-1}$. The sensitivity limit required to detect a single matched blazar is also the sensitivity limit required to detect a single matched FSRQ. Likewise, the sensitivity limit required to detect all matched FSRQs at all selected energies is the same at the limit required to detect all sources at all energies. The sensitivity limit required to detect a single BLL at any of the selected energies is 8.7 $\times$ $10^{-11}$ erg~cm$^{-2}$~s$^{-1}$, and the sensitivity limit required to detect all BLLs at all energies is 1.0 $\times$ $10^{-12}$ erg~cm$^{-2}$~s$^{-1}$.

\vspace{0.5em}

\noindent When a sensitivity limit of $10^{-10}$ erg~cm$^{-2}$~s$^{-1}$ is achieved, we can detect 7 of the cross-matched blazars at one or more of our selected energies: 3C 273, 3C 279, CTA 102, PKS 1830-211, 3C 454.3, PKS 1510-089, and 4C +71.07. These blazars are all FSRQs, and are also listed as among the brightest blazars at 1 MeV in Section \ref{sec:skymap}, and is roughly consistent with the detections made by COMPTEL \citep{Schoenfelder2000COMPTEL}. If we achieve a sensitivity limit of $10^{-11}$ erg~cm$^{-2}$~s$^{-1}$, we can detect 61 of the SED-modeled blazars at one or more of our selected energies. Of these, 50 are FSRQs, 8 are BLLs, and the other 3 are BCUs. If a sensitivity limit of $10^{-12}$ erg~cm$^{-2}$~s$^{-1}$ is achieved, we can detect all 76 of the SED-modeled blazars at one or more selected energies. Since the cross-matched blazar sample is intrinsically biased towards bright sources, with most hard X-ray and GeV $\gamma$-ray fluxes exceeding $10^{-12}$ erg~cm$^{-2}$~s$^{-1}$, it is expected that all SED-modeled blazars will be detectable when this sensitivity is achieved. Future MeV $\gamma$-ray instruments will have sensitivity limits in the $10^{-11}$ to $10^{-12}$ erg~cm$^{-2}$~s$^{-1}$ range, and thus these instruments will at baseline be able to detect a much larger blazar population than COMPTEL. The detectable counts of the cross-matched blazar population are shown by the step functions in Figure \ref{fig:cross_match_detection_counts}, and are broken down into their composite FSRQ and BLL counts. 

\vspace{0.5em}

\begin{figure*}[t!]
\centering

\includegraphics[width=0.99\textwidth]{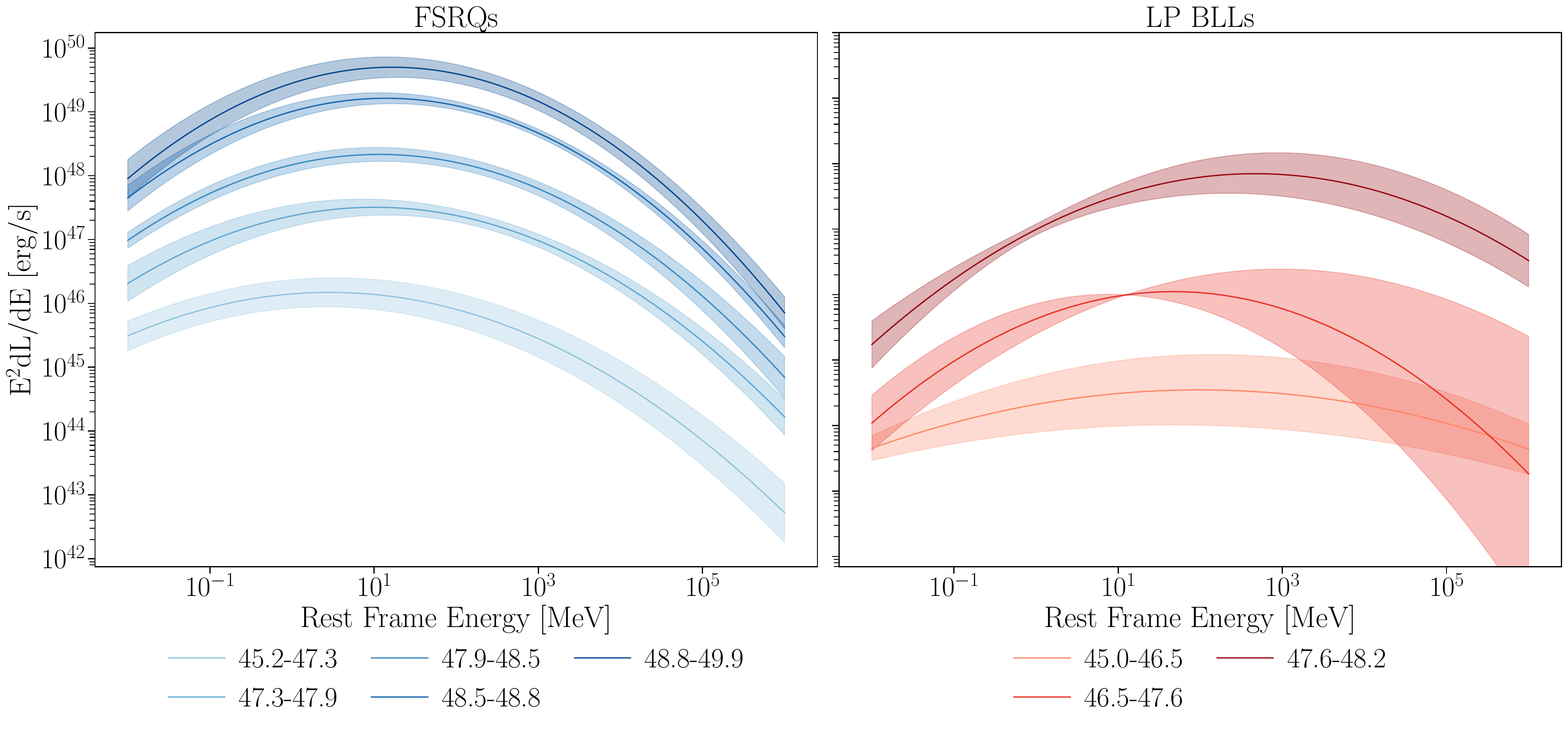}

\caption{Template SEDs for FSRQs (left), BLLs modeled with the LP (Right). Luminosity bin ranges are shown in logarithmic luminosity units [Log(erg~s$^{-1}$)] below each frame in the figure.}
\label{fig:template_SEDs}
\end{figure*}
 
\subsection{Cumulative Source Counts with Luminosity Functions}
\label{sec:lf_forecasting}

To make estimates of the total detectable source counts by future MeV instruments, we utilize luminosity functions (LFs) from published literature. In this work, we consider the luminosity-dependent density evolution (LDDE) form of blazar LFs as presented in \citet{Rajguru2025GammaRayLF} for Fermi-LAT $\gamma$-ray FSRQs, \citet{Toda2020FSRQsBAT} for Swift-BAT hard X-ray FSRQs, and \citet{Ajello_2014} for Fermi-LAT $\gamma$-ray BLLs. For \citet{Ajello_2014}, We select the $\mathrm{LDDE_2}$ as this model allowed different luminosity classes of BLLs to evolve differently, and was found to be the best-fitting model. We will henceforth refer to these works as R25, T20, and A13 respectively. For a complete review of the formalism for blazar luminosity functions, see Appendix \ref{sec:lf_review}.

\subsubsection{Template SEDs}
\label{sec:template_SEDS}

To determine the minimum luminosity source we can detect with a given instrument sensitivity, $L_{\gamma,min}(S,z)$, we construct averaged SEDs for FSRQs and LSP-type BLLs from their phenomenological SED models. We define the bolometric luminosity for each source as the luminosity in the 17 keV to 1 TeV energy range. For FSRQs, we divide the sources into five luminosity bins, and for LP-modeled BLLs, we divide the sources into three luminosity bins, with roughly the same number of sources for each blazar type. For luminosity-function calculations, we do not consider the spectra of cross-matched BCUs, as the luminosity functions adopted in this work were specifically derived for BLL and FSRQ populations separately. Within each bin we average the luminosity at each energy in the rest frame to construct the averaged SED. The averaged SED is then fit parametrically using the log parabola model for the FSRQs and LSP-type BLLs, as was conducted for the SED modeling in Section \ref{sec:phenom_model}. The template SEDs are shown in Figure \ref{fig:template_SEDs}. Error margins for the averaged SEDs are calculated using the jackknife technique, following the methodology of \citet{Ajello_2012}. We note that the errors for the LP BLLs are much larger due to small source counts per bin. All average FSRQ bins reach their peak energy between 2 and 20 MeV, and the LP BLL bins reach their peak between 40 and 500 MeV. While the LP-type BLLs have an overall lower-luminosity population and exhibit their peak at higher energies than the FSRQs, we do not observe the relationship between the peak energy and the bolometric luminosity predicted by the blazar sequence \citep{Fossati1998_sed, Ghisellini2017_FermiSequence} within each source population. 

\subsubsection{Cumulative Source Counts}
\label{sec:N_S_counts}

\begin{figure*}
\centering

\includegraphics[width=0.9\textwidth]{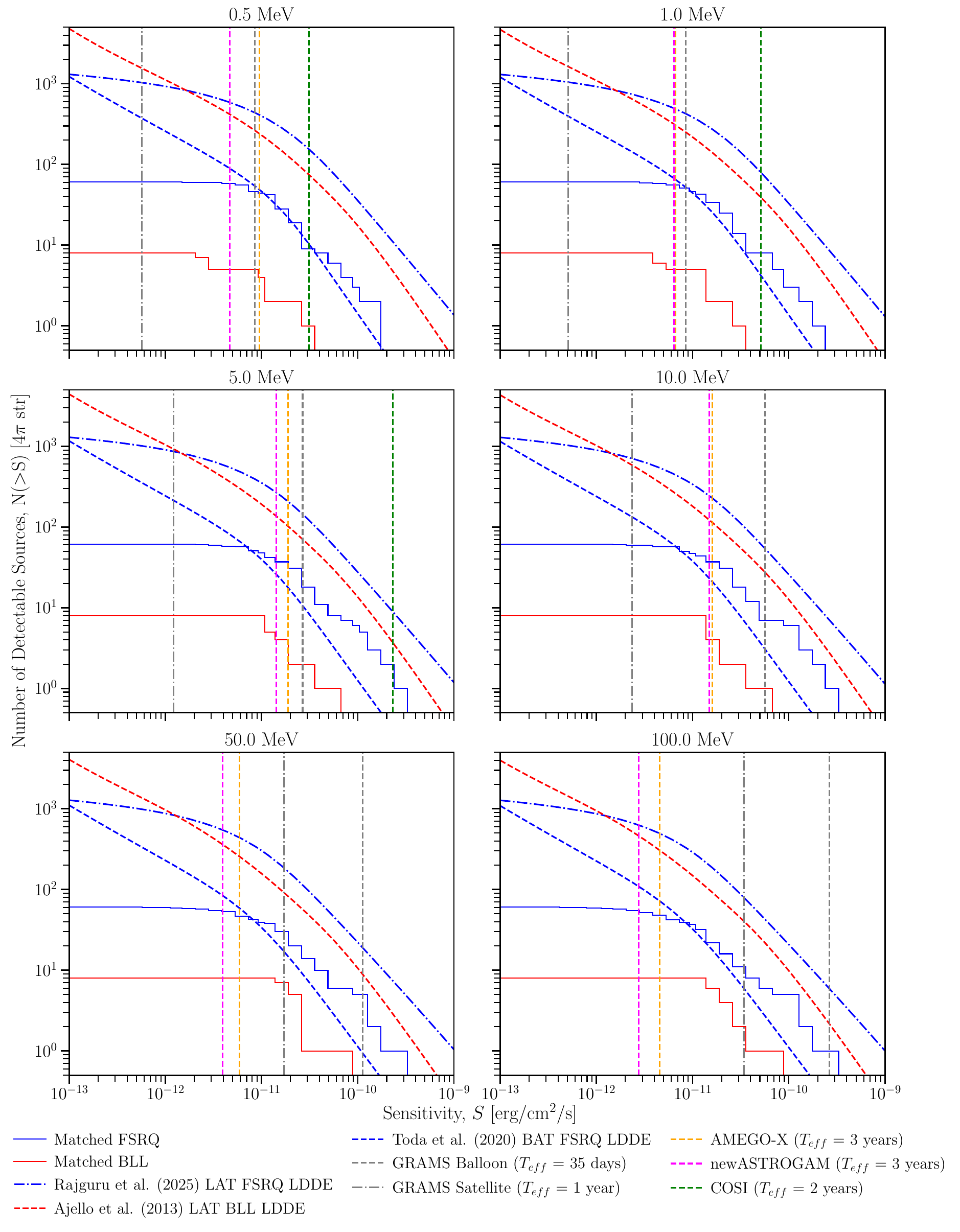}

\caption{$logN-logS$ plots at 0.5, 1, 5, 10, 50, and 100 MeV showing the detection counts of cross-matched blazars across a sensitivity range of $10^{-12}$ erg~cm$^{-2}$~s$^{-1}$ to $10^{-9}$ erg~cm$^{-2}$~s$^{-1}$. The total number of detectable blazars is shown by the solid black step function, with FSRQs and BL Lacs shown by the blue and red step functions, respectively. Dashed and dot-dashed continuum curves show the expected counts for best-fit LDDE LFs (see Section \ref{sec:N_S_counts} for handling of model uncertainty). Vertical lines show the continuum sensitivities of COSI (green), the GRAMS balloon (gray dashed), the GRAMS satellite (grey dot-dashed), AMEGO-X (orange), and newASTROGAM (magenta) at each respective energy.}
\label{fig:cross_match_detection_counts}
\end{figure*}

\noindent We determine the expected cumulative source counts $N(>$$S)$ from luminosity functions across the whole sky at a given sensitivity, $S$ using Equation \ref{eq:estimated_counts}. $L_{\gamma,min}(S,z)$ represents the luminosity of a source with a flux $S$ and a redshift $z$ at one of our selected energies. The remaining parameters $L_{\gamma,max}$, $z_{min}$, $z_{max}$, $\Gamma_{min}$, and $\Gamma_{max}$ are set based on the values used in each of the selected works.

\begin{multline}
      N(>S) = 4\pi \int_{\Gamma_{min}}^{\Gamma_{max}} \int_{z_{min}}^{z_{max}} \int_{L_{\gamma,min}(S,z)}^{L_{\gamma,max}} \\ \Phi(L_{\gamma}, V(z), \Gamma)\frac{dV}{dz}dL_{\gamma}dzd\Gamma
    \label{eq:estimated_counts}
\end{multline}

\noindent The lower limit of the luminosity integral is set by the sensitivity and the k-correction determined from the SED templates. For the remaining integrand limits, we adopt the limits presented in each of the selected works. For R25, $z_{min}=0.0001$, $z_{max}=5.0$, $\Gamma_{min}=1.0$, $\Gamma_{max}=4.0$, and $L_{\gamma,max}=7.3\times10^{48}$ erg/s. In T20, $z_{min}=0$, $z_{max}=6.0$, and $L_{\gamma,max}=10^{50}$ erg/s. In T20, the distribution of the photon index is not considered. Lastly, A13 uses the bounds $z_{min}=0.03$, $z_{max}=6.0$, $\Gamma_{min}=1.45$, $\Gamma_{max}=2.80$, and $L_{\gamma,max}=10^{52}$ erg/s. 

\vspace{0.5em}

\noindent We determine the cumulative source counts for sensitivities ranging from $10^{-12}$ erg~cm$^{-2}$~s$^{-1}$ to $10^{-9}$ erg~cm$^{-2}$~s$^{-1}$ for each luminosity function at each selected energy. The number of expected sources as a function of the sensitivity is shown in Figure \ref{fig:cross_match_detection_counts}, alongside the number of detectable cross-matched sources in each energy band. Here, we present the range of expected counts based on each luminosity function at $10^{-11}$ and $10^{-12}$ erg~cm$^{-2}$~s$^{-1}$ at each of our selected energies. For each luminosity function and sensitivity, the range of expected counts is reported as the minimum value to the maximum value for $N(>S)$ across the nine selected energies. We note that since the SED shape is known for only LP-modeled BLLs, our LF forecasts for BLLs only includes LP-modeled BLLs, and is would be larger when HSP-type BLLs are considered. When a sensitivity limit of $10^{-11}$ erg~cm$^{-2}$~s$^{-1}$ can be achieved, T20 estimates we can observe up to $\sim$30--50 FSRQs and R25 estimates we can observe up to $\sim$290--420 FSRQs. If a sensitivity of $10^{-12}$ erg~cm$^{-2}$~s$^{-1}$ can be achieved, T20 estimates we can observe up to $\sim$220--260 FSRQs and R25 estimates we can observe up to $\sim$870--940 FSRQs. The discrepancy between the expected number of detectable FSRQs estimated by T20 and R25 at both $10^{-11}$ and $10^{-12}$ erg~cm$^{-2}$~s$^{-1}$ is likely attributed to the difference in source population size from which each luminosity function was derived. R25 considers a source population consisting of 519 Fermi-LAT selected FSRQs, while T20 uses a much smaller sample of 53 Swift-BAT selected FSRQs. By utilizing these two luminosity functions from different blazar source populations, we set the uncertainty of our expected counts as the inter-model spread \citep{Inoue2015MeVAGN}. Using estimates from A13, we can expect a total population of roughly $\sim$150--250 BLLs when achieving a sensitivity of $10^{-11}$ erg~cm$^{-2}$~s$^{-1}$ and a total population of $\sim$950-1150 BLLs with a sensitivity limit of $10^{-12}$ erg~cm$^{-2}$~s$^{-1}$. We note that at a sensitivity limit of $10^{-12}$ erg~cm$^{-2}$~s$^{-1}$, source confusion due to poor MeV $\gamma$-ray instrument spatial resolution will become the primary limitation in resolving faint sources, as opposed to the instrument sensitivity.

\subsection{Forecast for Future MeV \texorpdfstring{$\gamma$}{gamma}-ray Instruments}
\label{sec:forecast}

Figure \ref{fig:cross_match_detection_counts} shows the combined forecast for next-generation MeV $\gamma$-ray instruments, utilizing the continuum sensitivity curves from the latest concepts of each mission in the literature. For GRAMS, we utilize the continuum sensitivity curves presented in \citet{Aramaki2020_GRAMS} for the GRAMS long-duration balloon flight ($T_{eff}=35$ days) and the GRAMS satellite with upgraded effective area ($T_{eff}=1$ year). For AMEGO, we utilize the sensitivity for the AMEGO-X mission presented in \citet{Perkins:2022amegox} ($T_{eff}=3$ years). For ASTROGAM, we utilize the sensitivity for newASTROGAM from \citet{Berge:2025newastrogam} ($T_{eff}=3$ years). For six of our selected energies (0.5, 1.0, 5.0, 10.0, 50.0, and 100 MeV), we show the number of detectable cross-matched blazars for sensitivities ranging from $10^{-9}$ erg~cm$^{-2}$~s$^{-1}$ to $10^{-12}$ erg~cm$^{-2}$~s$^{-1}$. In this forecast, we present the total number of cross-matched sources that are detectable by each instrument at one or more energies, and the maximum cumulative source count determined by each blazar luminosity function, $N(>S)_{max}$. We utilize this approach as the sensitivity of next-generation MeV $\gamma$-ray instruments varies by nearly an order of magnitude across different energies in the MeV band, and thus we seek to demonstrate the performance of each instrument at the energies it is most sensitive.

\vspace{0.5em}

\noindent In Figure \ref{fig:cross_match_detection_counts}, we also show the sensitivity for the upcoming COSI satellite ($T_{eff}=2$ years) \citep{Tomsick2023COSI} for the energies to which it is sensitive, up to 5 MeV. The continuum sensitivity of COSI is slightly improved from COMPTEL, ranging between $10^{-10}$ and $10^{-11}$ erg~cm$^{-2}$~s$^{-1}$. The COSI satellite will be capable of detecting 10 of the cross-matched blazars at 0.5 MeV, where COSI reaches its best continuum sensitivity of $3.1\times10^{-11}$ erg~cm$^{-2}$~s$^{-1}$. When considering the total population of expected blazars determined by luminosity functions, COSI could observe up to $\sim$170 FSRQs based on R25 estimates and $\sim$12 FSRQs based on T20 estimates. COSI could also expect a total observable population of $\sim$85 BLLs based on A13. With COSI planned to launch in 2027, the number of blazars which COSI detects during its operation will be critical to placing tighter constraints on the forecasts for future MeV $\gamma$-ray instruments. 

\vspace{0.5em}

\noindent With at least an order of magnitude improved sensitivity, next-generation MeV $\gamma$-ray instruments are expected to far surpass the capabilities of COMPTEL and COSI. With the GRAMS balloon, we can detect up to 60 of the 76 matched blazars and expect a total population of up to $\sim$530 FSRQs based on R25, $\sim$75 FSRQs based on T20, and $\sim$360 BLLs based on A13. The GRAMS instrument is most effective at low MeV energies, between 0.2 and 5 MeV, and will excel at crossing the lower end of the MeV gap. However, the sensitivity of the GRAMS instrument declines in the energy regime above 5 MeV. At 100 MeV, the GRAMS balloon is only capable of detecting one of the cross-matched blazars. At these energies, the total population that could be detected is also much more limited, with the GRAMS balloon only being capable of detecting $\sim$5 FSRQs based on R25, and no FSRQs or BLLs based on T20 and A13. Because of this, an instrument such as GRAMS will be complemented by concepts such as newASTROGAM and AMEGO. While these instruments do not achieve the same level of sensitivity at low MeV energies as GRAMS, they maintain sensitivity of roughly $10^{-11}$ to $10^{-12}$ erg~cm$^{-2}$~s$^{-1}$ across the MeV gap. With the sensitivity of newASTROGAM, we can detect 72 of the 76 matched blazars, and can expect a total population of $\sim$660 and $\sim$120 FSRQs from R25 and T20, and $\sim$515 BLLs from A13. Similarly, with AMEGO-X, we would be capable of detecting 69 of the 76 matched blazars at one or more energies. With the sensitivity of AMEGO-X, our luminosity functions demonstrate that we can expect a total population of $\sim$500 and $\sim$70 FSRQs from R25 and T20, and $\sim$325 BLLs from A13.

\vspace{0.5 em}

\noindent To achieve an expected blazar population similar to that of Fermi-LAT (on the order of $10^3$ sources) and detect all 76 SED modeled cross-matched blazars, a further sensitivity of $10^{-12}$ to $10^{-13}$ erg~cm$^{-2}$~s$^{-1}$ will be required. This level of sensitivity could be achieved with an instrument like that of the GRAMS satellite \citep{Aramaki2020_GRAMS}, with a much larger effective area and longer time of flight.

\section{High Redshift Detection Forecasting}
\label{sec:high_z_forecasting}

\begin{figure*}
    \centering
    \includegraphics[width=0.99\linewidth]{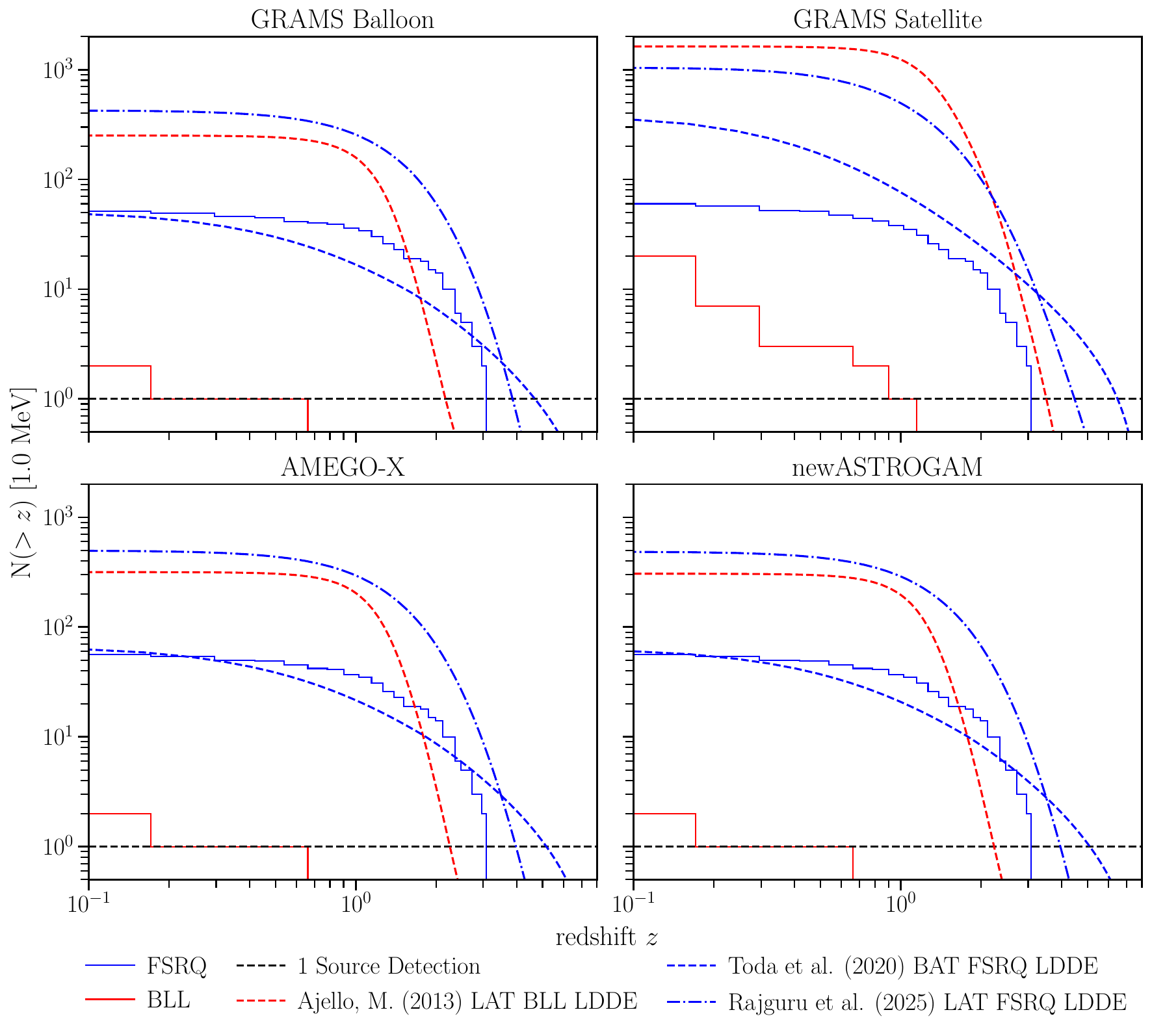}
    \caption{Cumulative redshift distribution for detectable cross-matched blazars and expected cumulative redshift distributions calculated from luminosity functions when assuming the sensitivity limits of the GRAMS balloon, GRAMS satellite, AMEGO-X, and newASTROGAM at 1 MeV.}
    \label{fig:N_z_counts}
\end{figure*}

Presently, the highest known redshift blazar is VLASS J041009.05-013919.88 ($z=7$), indicating the existence of jetted AGN in the epoch of reionization \citep{Banados2025}. At $\gamma$-ray energies, particularly in the 10 GeV to 10 TeV range, observation of blazars at high-redshifts is hindered by the attenuation of high-energy $\gamma$-rays by low-energy photons in the extragalactic background light \citep{Dwek2013}. At energies below $\sim$10--20 GeV, the attenuation of $\gamma$-rays by the extragalactic background light (EBL) is negligible up to $z>4$ \citep{Inoue2013EBL}. Thus, future MeV $\gamma$-ray instruments will be uniquely capable of expanding the redshift horizon for blazars in the $\gamma$-ray regime. Currently, hard X-ray instruments like Swift-BAT and NuSTAR have been ideal for searching for high-redshift MeV blazars \citep{Marcotulli_2020, Sbarrato_2013}, as it has been postulated that the most luminous blazars exhibit their high-energy peak in the MeV band \citep{Fossati1998_sed, Ghisellini2017_FermiSequence}, and MeV blazars are also bright in the hard X-ray band. As we discussed in Section \ref{sec:template_SEDS}, this correlation has yet to be proven, and our averaged SED construction demonstrated a lack of relationship. Thus, high-redshift blazar observations of the most luminous sources will be essential in resolving the validity of the blazar sequence. In addition, the observation of MeV blazars at high redshift will provide improved constraints on jet properties, as well as insight into the jet-accretion disk connection \citep{Marcotulli_2020, Paliya2020}.

\vspace{0.5em}

\noindent To forecast the redshift limits at which blazars could be detected by future MeV $\gamma$-ray instruments, we compute the cumulative redshift distribution $N(>z)$ as a function of redshift at a given energy for the cross-matched blazars and the expected blazar population from luminosity functions. For the cross-matched blazars, we count the number of blazars beyond a given redshift that we can detect based on their SED and the sensitivity limits of future MeV $\gamma$-ray instruments at selected energies, as discussed in Section \ref{sec:forecast}. To calculate the expected $N(>z)$ counts from luminosity functions at the same selected energies, we use Equation \ref{eq:N_z_calculation}. For this calculation, we utilize the same integration limits as discussed in Section \ref{sec:lf_forecasting}, with the exception of setting $z_{max}$ to be 8 for R25 and T20, given that FSRQs have been detected up to $z=7$ \citep{Banados2025}. This allows us to estimate the possible maximum redshift by extrapolating out to a higher-than-observed-redshift for FSRQs.

\begin{multline}
    N(z) = 4\pi\int_{z_{min}}^{z_{max}}\frac{dV}{d\Omega dz} \int_{\Gamma_{min}}^{\Gamma_{max}}\int_{L_{min}(S(E),z)}^{L_{max}} \\ \Phi(L_{\gamma}, V(z), \Gamma) dL d\Gamma 
    \label{eq:N_z_calculation}
\end{multline}

\noindent Figure \ref{fig:N_z_counts} shows both the $N(>z)$ counts for the cross-matched blazar population and the cumulative expected blazar population at 1 MeV for the sensitivity limits of the GRAMS balloon, GRAMS satellite, AMEGO-X, and newASTROGAM. The highest redshift blazar in the cross-matched population is PKS 0537-286 with a redshift of $z=3.104$. Our SED modeling of PKS 0537-286 estimates the flux of this source at 1 MeV to be 2.4$\times10^{-11}$ erg~cm$^{-2}$~s$^{-1}$, and thus this source will be detectable at 1 MeV by all four of the discussed future MeV $\gamma$-ray instruments. The cross-matched sample contains one other blazar at $z>3$, TXS 0800+618 ($z=3.033$). This source is slightly fainter than PKS 0537-286 with a flux of 8.7$\times10^{-12}$ erg~cm$^{-2}$~s$^{-1}$, and thus may be more challenging for instruments like the GRAMS balloon, newASTROGAM, and AMEGO-X. Both of these blazars are LSP-type FSRQs, with MeV $\gamma$-ray luminosities (0.1 MeV to 100 MeV) exceeding $10^{48}$ erg/s. In addition, both of these high-redshift sources reach the maximum of their high-energy component in the 1--10 MeV regime, thus making them strong candidates for future MeV $\gamma$-ray instrument observation. Sources observed by NuSTAR may also serve as ideal high-redshift observation candidates for future MeV $\gamma$-ray missions. SED modeling conducted for the six NuSTAR-observed FSRQs in \citet{Marcotulli_2020} indicates that the high-energy spectral peak occurs in the MeV energy regime, and that the flux at the spectral peak for all six sources lies in the $10^{-11}$ to $10^{-12}$ erg~cm$^{-2}$~s$^{-1}$ range. Thus, these sources may be bright enough to be detected by future MeV $\gamma$-ray instruments in the 0.1 to 10 MeV range. In addition, the highest redshift source detected by NuSTAR, B2 1023+25, also exhibits its spectral peak in the MeV regime, based on SED modeling done in \citet{Sbarrato_2013}. The flux of the peak determined in \citet{Sbarrato_2013} is estimated to be just below $10^{-12}$ erg~cm$^{-2}$~s$^{-1}$, and thus it may be too faint to be detected by instruments such as the GRAMS balloon, newASTROGAM, and AMEGO, but an improved mission such as the GRAMS satellite will be able to detect it. 

\vspace{0.5em}

\noindent We consider the expected maximum detectable redshift for each instrument at a given energy based on the estimates from luminosity functions to be the redshift at which $N(>z)\approx1$, where $N(>z)$ is an estimate of the total observable population at that redshift. We note that these estimates extrapolate beyond the sample population for which the luminosity function was derived. We report the maximum estimated detectable redshift as the largest $z_{max}$ value determined across the nine selected energies. Based on estimates of the cumulative redshift distribution for FSRQs utilizing the luminosity functions from R25 and T20, we can expect to detect up to $\sim10$ FSRQs at $z>3$ with instruments such as the GRAMS balloon, newASTROGAM, and AMEGO-X. For the GRAMS balloon, AMEGO-X, and newASTROGAM, we can expect blazars to be observable out to a maximum redshift of z $\approx$ 4, based on R25, and z $\approx$ 5--6 based on T20. Once the improved sensitivity of $10^{-12}$ to $10^{-13}$ erg~cm$^{-2}$~s$^{-1}$ is achieved with the GRAMS satellite, FSRQs could be detected out to z $\approx$ 6.5 (T20). The differences in the maximum redshift estimations between R25 and T20 can be attributed to the source population from which the LDDE LF parameters were derived. The sample used for T20 contains FSRQs up to $z$ = 4.715 \citep{Toda2020FSRQsBAT}, while R25 contains FSRQs only out to $z$ = 3.59 \citep{Rajguru2025GammaRayLF}. In addition, we can expect a hard X-ray LF to estimate higher-redshift sources, given that the most luminous FSRQs are brightest in the hard X-ray band, as discussed above, as well as the limitation of high-redshift observation that GeV and TeV instruments face due to the EBL. 

\vspace{0.5em}

\noindent While BLLs have primarily been observed at low redshifts, they have been reported out to a redshift of $z$ = 3.5 by Fermi-LAT \citep{Ajello_2022}, and a potential BLL candidate has been reported at a redshift as high as $z$ = 6.57 \citep{Frey2024J2331}. The highest-redshift BLL in the cross-matched sample that can be detected by the GRAMS balloon, AMEGO-X, and newASTROGAM at 1 MeV is S5 1803+784, at a redshift of $z$ = 0.68 and a flux of 9.6$\times10^{-12}$ erg~cm$^{-2}$~s$^{-1}$. LSP-type BLLs have been notably observed out to higher redshift than their HSP counterparts, given that they belong to a higher-luminosity class of BLLs. For the GRAMS balloon, the maximum expected detectable redshift for BLLs based on A13 is z $\approx$ 2.4. AMEGO-X and newASTROGAM may be able to detect slightly higher-redshift BLLs at z $\approx$ 2.5 and z $\approx$ 2.8, respectively. With the GRAMS satellite, BLLs could be detected as distant as z $\approx$ 3.5.

\section{Discussion}
\label{sec:discussion}

\subsection{Cross-Match Forecasting}
\label{sec:cross_match_discussion}

\begin{figure}[h]
    \centering
    \includegraphics[width=\linewidth]{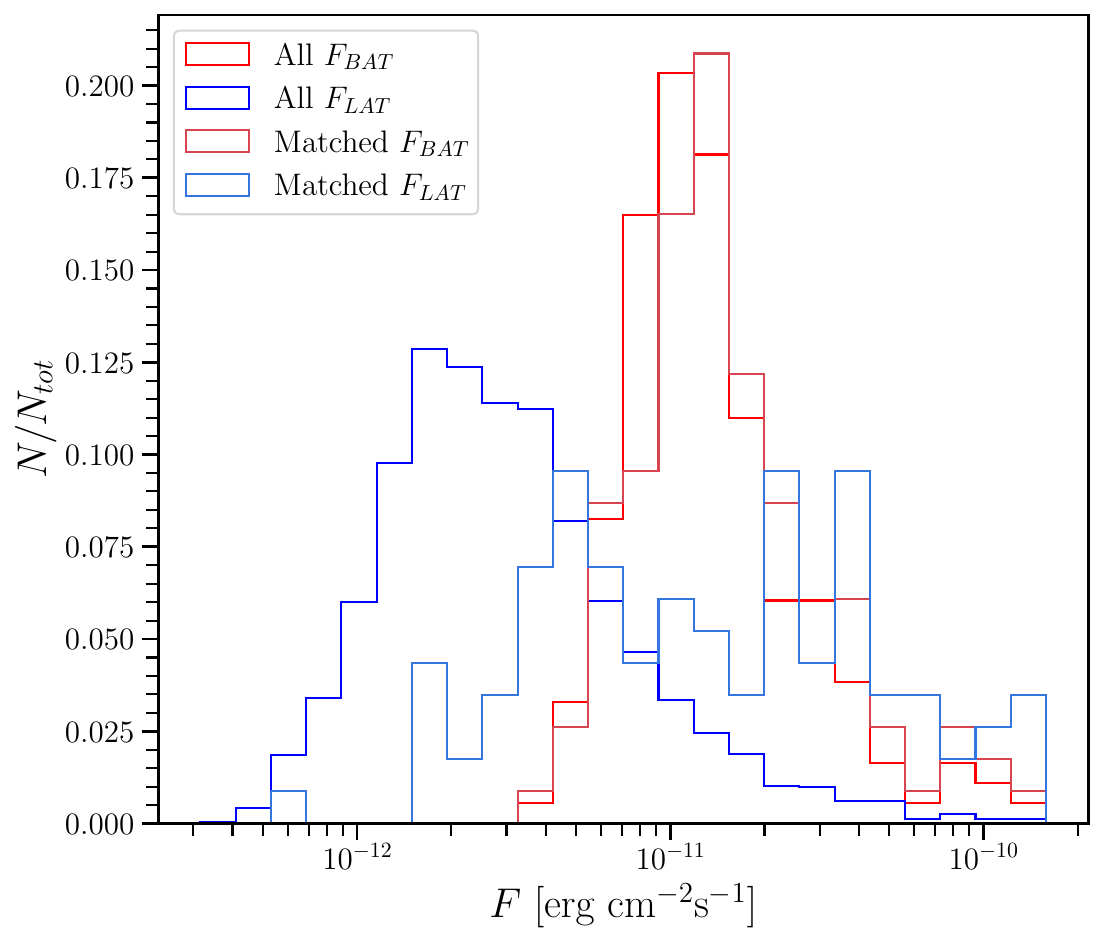}
    \caption{Histogram for the percentages of energy fluxes in the BAT (14 keV--195 keV) and LAT (100 MeV--100 GeV) bands of the cross-matched blazar sample compared to the total BAT and LAT blazar populations.}
    \label{fig:BAT_LAT_flux_pop}
\end{figure}

By cross-matching the latest data releases of the Swift-BAT and Fermi-LAT catalogs, we identified a sample of 115 blazars with both hard X-ray and GeV $\gamma$-ray coverage that are promising targets for future MeV $\gamma$-ray instruments. For these blazars, we are able to make more constrained flux estimates than we could by extrapolating across the MeV gap using only LAT or BAT data. However, the cross-matched blazar sample is still limited, largely because Swift-BAT has thus far only detected approximately 200 blazars in total, compared to the nearly 4000 detected by Fermi-LAT. The majority of Swift-BAT blazars have hard X-ray fluxes exceeding $10^{-12}$ erg~cm$^{-2}$~s$^{-1}$ \citep{Lien_2025}, thus roughly setting the sample's flux limit. This limit is consistent with our finding that most cross-matched sources are detectable when a sensitivity limit of $10^{-12}$ erg~cm$^{-2}$~s$^{-1}$ is achieved. Figure \ref{fig:BAT_LAT_flux_pop} compares the distribution of integrated energy fluxes for the full BAT and LAT blazar populations to that of the cross-matched sample. The LAT blazar population contains a larger percentage of faint sources than the cross-matched LAT blazars. If a larger sample of fainter sources were to be detected by a hard X-ray instrument, forecast numbers could theoretically close the gap between luminosity function estimates and cross-match-sample predictions. 

\vspace{0.5 em}

\noindent This same selection effect is also reflected in the distribution of cross-matched blazar source types. The relative population sizes of FSRQs and BLLs in the cross-matched blazar sample are more similar to those of the BAT population than the LAT population. In LAT, the ratio of FSRQs to BLLs is roughly two to three, whereas in the cross-matched population the ratio is roughly three to two. The Second Fermi Low Energy (2FLE) Catalog \citep{FermiLAT2026_2FLE}, which contains sources detected by Fermi-LAT in the 20--200 MeV range, is more consistent with the populations of cross-matched and BAT blazars than with the LAT blazars. The 2FLE catalog contains 218 FSRQs, 67 BLLs, and 29 BCU thus representing a roughly three to one ratio between FSRQs and BLLs. This similarity between the 2FLE blazars and the cross-matched blazars provides further validation for the distribution of blazars that will be detected by future MeV instruments. When comparing the ratios of SED classification between the cross-matched blazars and the LAC blazars, LSP- and HSP-type blazars are the dominant populations in the cross-matched sample, composing 93$\%$ of all matched blazars. The primary discrepancy between the cross-matched blazars and the LAC blazars is the percentage of blazars with unknown SED classification. Since the cross-matched sample contains mostly bright sources with broadband observational coverage, the peak of the synchrotron component can be determined, unlike for faint LAC sources. ISP-type blazars, which are in a roughly one-to-one ratio with HSP-type blazars in the LAC sample, are mostly absent from the cross-matched sample. In the broadband SED of ISP-type blazars, the valley between the synchrotron and inverse-Compton components lies around the hard X-ray band, and thus these sources may be too faint in the hard X-ray to be detected by Swift-BAT. Provided that the rising edge of the high-energy component will be situated in the MeV band for ISP-type blazars, it is likely that this class of blazars will be detectable by future MeV instruments and may be undercounted in our forecast estimates.

\vspace{0.5 em}

\noindent This sample limitation may also provide justification for the lack of the blazar-sequence anti-correlation found in the averaged SEDs for FSRQs in Section \ref{sec:template_SEDS}. If lower luminosity FSRQs are more likely to achieve their spectral peak at high MeV or low GeV energies, their hard X-ray flux may be below the limit needed to be detected by Swift-BAT. Thus, our sample may be biased by only containing low-luminosity FSRQs that peak in the MeV range, which represent a minority of the low luminosity FSRQ population.  

\subsection{Luminosity Function Forecasting}

We also used blazar luminosity functions to forecast the cumulative source counts that would be expected for a given instrument sensitivity, which serve as an upper limit for detectability. This calculation for the MeV band has already been completed in works such as \citet{Inoue2015MeVAGN} and \citet{Marcotulli:2022BASS33}. These works calculate the counts in the integrated bands, 0.3--1, 1--3, 3--10, 10--30, and 30--100 MeV, and do not compare directly to specific instrument sensitivities. As such, we cannot directly compare these band calculations to the specific energy calculations in this work. However, we can compare the ranges of estimates for bands and energies. \citet{Inoue2015MeVAGN} adopts the modified pure luminosity evolution (PLE) luminosity function presented in \citet{Ajello_2009} (A09) for Swift-BAT FSRQs and the LDDE luminosity function in \citet{Ajello_2012} (A12) for Fermi-LAT FSRQs. \citet{Inoue2015MeVAGN} estimates that when a sensitivity limit of $10^{-11}$ erg~cm$^{-2}$~s$^{-1}$ is achieved, we can expect $\sim$5--90 FSRQs based on A09, and $\sim$220--380 FSRQs based on A12. These estimates are roughly comparable to those determined from the luminosity functions for FSRQs we adopt in this work, where T20 estimated $\sim$30--50 FSRQs and R25 estimated $\sim$290--420 FSRQs. With a sensitivity limit of $10^{-12}$ erg~cm$^{-2}$~s$^{-1}$, \citet{Inoue2015MeVAGN} reports that we can expect $\sim10^3$--$10^4$ FSRQs based off A09, and with A12, we can expect $\sim10^3$ FSRQs. At this sensitivity, the expected FSRQ population based on the $\gamma$-ray luminosity functions in A12 and R25 is in relatively good agreement, with R25 expecting $\sim$870--940 FSRQs. The use of different luminosity function models between A09 and T20 can explain the discrepancy between the Swift-BAT based expected counts for FSRQs. The LDDE luminosity function used in T20 predicts that the space density of FSRQs will decrease beyond the critical redshift, whereas the PLE model used in A09 predicts that the space density remains constant, and thus there could be a much larger population of FSRQs even out at high redshifts. Similarly, \citet{Marcotulli:2022BASS33} shows that the PLE model for BAT FSRQs is preferred, and also estimates $\sim10^3$--$10^4$ FSRQs with a sensitivity of $10^{-12}$ erg~cm$^{-2}$~s$^{-1}$. At $10^{-11}$ erg~cm$^{-2}$~s$^{-1}$, \citet{Marcotulli:2022BASS33} predicts a slightly higher number of FSRQs than A09 and T20, with an expected cumulative population of $\sim30$--300 FSRQs. 

\vspace{0.5 em}

\noindent As mentioned in Section \ref{sec:N_S_counts}, an important limitation to estimating source counts beyond $10^3$ sources is the poor spatial resolution of future MeV $\gamma$-ray instruments. The spatial resolution of MeV instruments is roughly 3--$5^{\circ}$ \citep{Aramaki2020_GRAMS, Perkins:2022amegox, Tomsick2023COSI, Berge:2025newastrogam}. If $10^3$ sources are uniformly distributed across the sky, the average angular distance between sources will be roughly $3.2^\circ$, making it difficult to resolve faint sources. It is likely that MeV spatial resolution will become a limitation before these counts are reached, as we have discussed in Section \ref{sec:skymap}.
 
\subsection{MeV \texorpdfstring{$\gamma$}{gamma}-ray Neutrino Connection}

In observing blazars in the MeV $\gamma$-ray regime, we may also gain critical insight into the connection between potential neutrino emission from blazars and associated MeV $\gamma$-rays, thus making way for multimessenger studies with future MeV $\gamma$-ray instruments. Famously, the $\gamma$-ray emitting blazar TXS 0506+056 was strongly associated ($\sim3\sigma$) with the IceCube- 170922A neutrino event in 2017, with a follow up investigation determining further neutrino coincident events at $3.5\sigma$ in 2014 and 2015 \citep{IceCube:2018dnn, IceCube:2018cha}. These studies of TXS 0506+056 have offered the strongest evidence thus far that blazars are candidate sources astrophysical neutrinos. 

\vspace{0.5 em}

\noindent While TXS 0506+056 has been observed by Fermi-LAT, it has not been observed by Swift-BAT, and thus remains unmatched in out sample. \citet{Petropoulou_2020} and \citet{Acciari_2022} conduct broadband SED modeling of TXS 0506+056 using data from Swift-XRT, NuSTAR, and Fermi-LAT $\gamma$-ray in various observation epochs. In all cases, the high-energy peak for TXS 0506+056 is shown to occur around 1 GeV, with the rising edge of the high-energy component situated in the MeV regime. \citet{Petropoulou_2020} shows the minima of the valley between the synchrotron component and the high-energy component occurring around 10 keV, with flux in all epochs of less than $10^{-12}$ erg~cm$^{-2}$~s$^{-1}$. The low hard X-ray flux of TXS 0506+056 may explain why it remains undetected by Swift-BAT. TXS 0506+056 is notably also an ISP type blazar, as we discussed in Section \ref{sec:cross_match_discussion}, may belong to a class of underrepresented sources in the cross-matched blazar sample. From there the flux rises to the high energy peak at around $10^{-11}$ erg~cm$^{-2}$~s$^{-1}$. In the observations of \citet{Acciari_2022}, the flux of the valley minima is around $10^{-12}$ erg~cm$^{-2}$~s$^{-1}$, rising to its peak at nearly $10^{-10}$ erg~cm$^{-2}$~s$^{-1}$. Provided these modeled flux estimates, TXS 0506+056 could very likely be detectable by future MeV $\gamma$-ray instruments. If a distinct emission component or polarization can be detected from TXS 0506+056 in coincidence with a neutrino alert, such a detection would provide strong evidence for hadronic processes in blazar jets. 

\vspace{0.5 em}

\noindent Similar studies could be conducted on observations of the FSRQ PKS 1502+106, which was spatially coincident with the neutrino event IceCube-190730A. PKS 1502+106 is one of the brightest $\gamma$-ray blazars, and has been observed in both quiescent and flaring epochs. \citet{Rodrigues_2021} applies both lepto-hadronic and proton-synchrotron emission models to the quiescent SED of PKS 1502+106, as well as its SED during hard and soft $\gamma$-ray flaring states. During the quiescent state, the lepto-hadronic model predicts an flux in the MeV band exceeding $10^{-11}$ erg~cm$^{-2}$~s$^{-1}$, while the proton-synchrotron model predicts flux in the MeV band between $10^{-12}$ erg~cm$^{-2}$~s$^{-1}$ and $10^{-11}$ erg~cm$^{-2}$~s$^{-1}$. Given these flux estimates, PKS 1502+106 could be detected by instruments such as the GRAMS balloon, newASTROGAM, and AMEGO-X during quiescent states, but would much more likely need an instrument as the sensitivity level of the GRAMS satellite. During $\gamma$-ray flaring states, PKS 1502+106 would certainly be detectable by future MeV $\gamma$-ray instruments, with both models during the soft $\gamma$-ray flaring state estimating flux in the MeV band between $10^{-11}$ erg~cm$^{-2}$~s$^{-1}$ and $10^{-10}$ erg~cm$^{-2}$~s$^{-1}$. During hard $\gamma$-ray flaring states, both models predict the flux in the MeV band could even exceed $10^{-10}$ erg~cm$^{-2}$~s$^{-1}$, making this source an ideal target for next generation $\gamma$-ray instruments.

\vspace{0.5em}

\noindent Hadronic emission components associated with neutrino production are an additional element of uncertainty that should be considered in our forecast. If the hadronic emission contributes significantly to the SED, our forecast may be underestimating the expected counts. The potential emergence of emission from hadronic components in the MeV regime is another element of uncertainty for sources where the MeV regime is situated between the primary non-thermal emission components. If emission associated with hadronic processes contributes significantly to the SED of these types of sources, then their spectral shape may be more complicated than a simple SBPL model could predict, as discussed in Section \ref{sec:phenom_model}.

\section{Conclusions}

In this work, we presented a forecast for future MeV $\gamma$-ray instruments utilizing a population of cross-matched blazars in conjunction with cumulative expected blazar populations based on luminosity functions. The results of this work are summarized in the following points:

\begin{enumerate}

\item We identify a sample of 115 firmly matched blazars through matching the the fourth data release of the fourth Fermi-LAT source catalog and Swift-BAT 157-Month source catalog using the methodology presented in \citet{Tsuji_2021}. The firmly matched blazar sample contains $\sim$20$\%$ more firmly matched blazars than presented in \citet{Tsuji_2021}, composed of mostly  FSRQs (61) and HSP-type BLLs (34), with a smaller population of LSP-type BLLs (7). 

\item We use the multiwavelength time averaged spectral data from LAT and BAT to extrapolate the SED of 76 of the 115 matched sources. These sources are selected as those with hard BAT and soft LAT spectral indices, and thus we can model flux extimates in the MeV band with a simple log parabola. For matched blazars with soft BAT and hard LAT spectral indices, we do not use modeled flux estimates, as the uncertainty of the flux for these sources could vary by orders of magnitude based on the spectral shape between the non-thermal SED components.

\item Using selected SED models, we make model-driven flux estimates at nine selected energies. We show the extragalactic sky for SED-modeled blazars in our sample at 1 MeV, and discuss the brightest matched-sample blazars in the MeV regime as ideal targets for early MeV observations. We also demonstrate the counts for the cross-matched source population that can be detected when different orders of magnitude sensitivity can be achieved. We find that 7, 61, and 76 of the 76 cross-matched blazars can be detected at $10^{-10}$, $10^{-11}$, and $10^{-12}$ erg~cm$^{-2}$~s$^{-1}$, respectively. 

\item We pair source count estimates from the cross-matched sample with cumulative source count estimates in the MeV regime based on luminosity functions. In doing so, we provide a data-driven lower limit using SED flux modeling of cross-matched blazars to the detection capabilities of future MeV instruments that can be paired with the upper limits posited by luminosity function based studies. We utilize the luminosity functions derived in \citet{Rajguru2025GammaRayLF} for LAT FSRQs, \citet{Toda2020FSRQsBAT} for BAT FSRQs, and \citet{Ajello_2014} for LAT BLLs to make estimates on the total cumulative population that can be expected for a given sensitivity of instrument. With a sensitivity limit $10^{-11}$ erg~cm$^{-2}$~s$^{-1}$, the total FSRQ population is expected to range from a few tens of FSRQs \citep{Toda2020FSRQsBAT} to a few hundreds \citep{Rajguru2025GammaRayLF}, and the total BLL population is expected to also be a few hundreds \citep{Ajello_2014}. With a sensitivity of $10^{-12}$ erg~cm$^{-2}$~s$^{-1}$, we can expect a total population of a few hundred FSRQs \citep{Toda2020FSRQsBAT} to approximately 1000 FSRQs \citep{Rajguru2025GammaRayLF}, and a total population of roughly 1000 BLLs. 

\item We then utilize generalized count estimates for the cross-matched population and cumulative source populations to make a forecast for the capabilities for future MeV instruments. We demonstrate based on our counts estimates that next-generation instruments like the GRAMS balloon, newASTROGAM, and AMEGO-X missions will surpass the capabilities of COMPTEL, as well as current generation instruments like COSI. Next generation instruments with detector upgrades like the GRAMS satellite will have excellent capabilities with order magnitude improved sensitivity from other next generation instruments. 

\item We also demonstrate the capability of future MeV $\gamma$-ray instruments to observe high-redshift sources, as MeV $\gamma$-ray observation is not hindered by EBL attenuation like GeV $\gamma$-ray instruments are. With next generation instruments we will be able to observe cross-matched sources like PKS 0537-286 and TXS 0800+618 at $z>3$, and potentially observe even more distant FSRQs up to $z$ $\approx$ 6 and BLLs up to $z$ $\approx$ 2.8. 

\item Lastly, we discuss the detectability of blazars that have been proposed to be neutrino emitters, based on their spacial coincidence with IceCube events. We compare flux levels estimated from SED models in the literature for TXS 0506+056 and PKS 1502+106 to demonstrate that they will be detectable by future MeV $\gamma$-ray instruments to support multimessenger studies.

\end{enumerate}

\section{Acknowledgments}

This work was supported by funding for the GRAMS collaboration through the NASA APRA grant, No.22-APRA22-0128 (80NSSC23K1661), and the Alfred P. Sloan Foundation in the US, as well as the Japan Society for the Promotion of Science (JSPS) in Japan. In addition, this work was supported by the Northeastern University Undergraduate Research and Fellowships office by funding through the PEAK Ascent award (Summer 2024), PEAK Summit awards (Fall 2024, Spring 2025), and the AJC Merit Scholarship. The author would also like to acknowledge the support and mentorship of his fellow co-authors through developing and executing this project. 

\vspace{0.5 em}

\noindent Y.I. is supported by JST FOREST Program (Grant Number JPMJFR256L, Japan), MEXT Supporting Pioneering Research through AI for 1,000 Discovery challenges Program (SPReAD) Japan Grant Number JPMXP1726291302, and JSPS KAKENHI grant Nos. JP22K18277, JP26H00604, and JP26K22331.

\vspace{0.5 em}

\noindent This work made use of Gammapy \citep{gammapy:2023}, a community-developed Python package. The Gammapy team acknowledges all Gammapy past and current contributors, as well as all contributors of the main Gammapy dependency libraries: \href{https://numpy.org/}{NumPy}, \href{https://scipy.org/}{SciPy}, \href{http://www.astropy.org}{Astropy}, \href{https://astropy-regions.readthedocs.io/}{Astropy Regions}, \href{https://scikit-hep.org/iminuit/}{iminuit}, \href{https://matplotlib.org/}{Matplotlib}.

\clearpage

\appendix

\section{Catalog of Matched Blazars as MeV Observation Candidates}
\label{sec:MeV_Blazar_catalog}

\vspace{0.5cm}
\begin{longtable*}{lcccccc}
\caption{Firmly Matched Blazars\label{tab:MeV_Blazars}}\\
\hline\hline
Fermi Name & Fermi Class & Swift Name & Swift Class & Redshift\tablenotemark{a} & SED Type \\
\hline
\endfirsthead
\multicolumn{6}{c}%
{{\tablename\ \thetable{} -- continued}} \\
\hline\hline
Fermi Name & Fermi Class & Swift Name & Swift Class & Redshift & SED Type \\
\hline
\endhead
\hline \multicolumn{6}{r}{{Continued on next page}} \\
\endfoot
\hline
\endlastfoot
1ES 0033+595 & bll & 1ES 0033+595 & BZB & 0.086 & HSP \\
1ES 0120+340 & bll & SHBL J012308.7+342049 & BZB & 0.272 & HSP \\
B3 0133+388 & bll & B3 0133+388 & BZB & None & HSP \\
TXS 0134+579 & bll & 2FHL J0138.2+5811 & Beamed AGN & None & HSP \\
1RXS J015658.6-530208 & bll & RBS 259 & Beamed AGN & None & HSP \\
PKS 0202-17 & fsrq & QSO B0202-172 & FSRQ & 1.7395 & LSP \\
TXS 0210+515 & bll & 2MASX J02141794+5144520 & BZG & 0.049 & HSP \\
S5 0212+73 & fsrq & [HB89] 0212+735 & BZQ & 2.367 & LSP \\
1ES 0229+200 & bll & QSO B0229+200 & BZG & 0.1396 & HSP \\
RBS 0351 & bll & BZB J0244-5819 & Beamed AGN & 0.2645 & HSP \\
4C +33.06 & bcu & 4C +33.06 & Beamed AGN & None & LSP \\
1RXS J032521.8-563543 & bll & 2MASX J03252346-5635443 & BZG & 0.0602 & HSP \\
TXS 0322+222 & fsrq & QSO J0325+2224 & FSRQ & 2.066 & LSP \\
1ES 0347-121 & bll & QSO B0347-121 & BZG & 0.18 & HSP \\
PKS 0352-686 & bll & PKS 0352-686 & BZB & 0.087 & HSP \\
NRAO 150 & fsrq & 4C +50.11 & BZQ & 1.52 & LSP \\
PKS 0402-362 & FSRQ & PKS 0402-362 & BZQ & 1.417 & LSP \\
PKS 0403-13 & fsrq & [HB89] 0403-132 & BZQ & 0.5706 & LSP \\
PKS 0454-46 & fsrq & PKS 0454-46 & FSRQ & 0.85 & LSP \\
1ES 0502+675 & bll & 87GB 050246.4+673341 & BZB & 0.314 & HSP \\
PKS 0528+134 & FSRQ & PKS 0528+134 & BZQ & 2.06 & LSP \\
PKS 0537-441 & BLL & [HB89] 0537-441 & Beamed AGN & 0.894 & LSP \\
PKS 0537-286 & fsrq & [HB89] 0537-286 & BZQ & 3.104 & LSP \\
PKS 0548-322 & bll & PKS 0548-322 & BZB & 0.069 & HSP \\
B2 0552+39A & fsrq & [HB89] 0552+398 & BZQ & 2.365 & LSP \\
PKS 0637-75 & fsrq & PKS 0637-752 & BZQ & 0.651 & LSP \\
TXS 0637-128 & bll & PMN J0640-1253 & Beamed AGN & 0.135 & HSP \\
PMN J0641-0320 & FSRQ & PMN J0641-0320 & Beamed AGN & 1.196 & LSP \\
RX J0648.7+1516 & bll & RX J0648.7+1516 & Beamed AGN & 0.179 & HSP \\
1ES 0647+250 & bll & QSO B0647+250 & Beamed AGN & 0.203 & HSP \\
PKS 0706-15 & bll & PKS 0706-15 & BZB & 0.142 & HSP \\
1H 0658+595 & bll & 2MASX J07103005+5908202 & BZB & 0.125 & HSP \\
S5 0716+71 & BLL & [HB89] 0716+714 & Beamed AGN & 0.3 & ISP \\
PKS 0723-008 & bcu & PKS 0723-008 & BZQ & 0.128 & LSP \\
NVSS J073326+515355 & bcu & 2MASX J07332681+5153560 & BZG & 0.065 & HSP \\
PKS 0736+01 & fsrq & QSO B0736+01 & FSRQ & 0.19 & LSP \\
B2 0743+25 & fsrq & B2 0743+25 & BZQ & 2.9793 & LSP \\
TXS 0800+618 & fsrq & CGRaBS J0805+6144 & BZQ & 3.033 & LSP \\
SUMSS J082627-640414 & bll & 4FGL J0826.4-6404 & Beamed AGN & 0.24 & HSP \\
4C +71.07 & FSRQ & [HB89] 0836+710 & BZQ & 2.172 & LSP \\
1ES 0927+500 & bll & 2MASS J09303759+4950256 & Beamed AGN & 0.1867 & HSP \\
RXC J0934.4-1721 & bll & 2MASS J09343014-1721215 & Beamed AGN & 0.2499 & ISP \\
PMN J0957-1350 & fsrq & ICRF J095718.1-135001 & FSRQ & 1.323 & LSP \\
1ES 1028+511 & bll & 2MASX J10311847+5053358 & BZB & 0.3604 & HSP \\
S5 1039+81 & fsrq & [HB89] 1039+811 & Beamed AGN & 1.26 & LSP \\
1ES 1101-232 & bll & 2MASX J11033765-2329307 & BZB & 0.186 & HSP \\
Mkn 421 & BLL & Mrk 421 & BZB & 0.03 & HSP \\
PKS 1127-14 & fsrq & PKS 1127-14 & BZQ & 1.184 & LSP \\
RX J1136.5+6737 & bll & 2MASX J11363009+6737042 & BZG & 0.1342 & HSP \\
PKS 1143-696 & fsrq & PKS 1143-696 & BZQ & 0.244 & LSP \\
B2 1150+33A & fsrq & 7C 1150+3324 & Beamed AGN & 1.3976 & LSP \\
4C +49.22 & FSRQ & 4C +49.22 & BZQ & 0.334 & LSP \\
Ton 599 & fsrq & FBQS J1159+2914 & Beamed AGN & 0.7247 & LSP \\
PG 1218+304 & bll & FBQS J1221+3010 & BZB & 0.1836 & HSP \\
4C +04.42 & fsrq & 4C +04.42 & BZQ & 0.965 & LSP \\
4C +21.35 & FSRQ & PG 1222+216 & BZQ & 0.432 & LSP \\
3C 273 & FSRQ & 3C 273 & BZQ & 0.1583 & LSP \\
3C 279 & FSRQ & 3C 279 & BZQ & 0.5362 & LSP \\
PMN J1310-5552 & bcu & IGR J13109-5552 & BZQ & 1.5591 & HSP \\
PKS 1329-049 & fsrq & PKS 1329-049 & BZQ & 2.15 & LSP \\
PKS 1335-127 & fsrq & [HB89] 1334-127 & BZQ & 0.539 & LSP \\
4C +19.44 & fsrq & [HB89] 1354+195 & Beamed AGN & 0.72 & None \\
H 1426+428 & bll & 1ES 1426+428 & BZB & 0.129 & HSP \\
PMN J1508-4953 & bcu & PMN J1508-4953 & BZQ & None & LSP \\
PKS 1510-089 & FSRQ & PKS 1510-08 & BZQ & 0.36 & LSP \\
AP Librae & bll & AP Lib & Beamed AGN & 0.049 & LSP \\
PG 1553+113 & BLL & QSO B1553+113 & Beamed AGN & 0.36 & HSP \\
PKS B1622-297 & FSRQ & PKS 1622-29 & BZQ & 0.815 & LSP \\
4C +38.41 & FSRQ & 4C 38.41 & FSRQ & 1.814 & LSP \\
3C 345 & FSRQ & 3C 345 & BZQ & 0.5928 & LSP \\
Mkn 501 & BLL & Mrk 501 & BZB & 0.0337 & HSP \\
2MASS J16561677-3302127 & fsrq & 2MASS J16561677-3302127 & BZQ & 2.4 & LSP \\
S5 1803+784 & bll & [HB89] 1803+784 & Beamed AGN & 0.68 & LSP \\
PKS 1824-582 & fsrq & 2MASS J18291242-5813554 & FSRQ & 1.531 & None \\
PKS 1830-211 & FSRQ & PKS 1830-21 & BZQ/Lense & 2.5 & LSP \\
PKS B1921-293 & fsrq & [HB89] 1921-293 & Beamed AGN & 0.3526 & LSP \\
PKS 1936-623 & bll & PKS 1936-623 & Beamed AGN & None & LSP \\
MG2 J194359+2118 & bcu & SWIFT J1943536.21+211822.9 & BZB & None & None \\
1ES 1959+650 & bll & QSO B1959+650 & BZB & 0.047 & HSP \\
PKS 2005-489 & BLL & PKS 2005-489 & BZB & 0.071 & HSP \\
MG2 J201534+3710 & FSRQ & QSO B2013+370 & BZQ & 0.859 & LSP \\
RGB J2056+496 & bcu & RX J2056.6+4940 & Beamed AGN & None & HSP \\
PKS 2142-75 & FSRQ & [HB89] 2142-758 & Beamed AGN & 1.139 & LSP \\
PKS 2149-306 & fsrq & PKS 2149-306 & BZQ & 2.345 & LSP \\
NRAO 676 & fsrq & 87GB 215950.2+503417 & BZQ & 1.899 & LSP \\
BL Lac & BLL & BL Lac & BZB & 0.0686 & LSP \\
PKS 2223+21 & fsrq & QSO J2225+2118 & FSRQ & 1.9593 & LSP \\
PKS 2227-08 & FSRQ & PKS 2227-088 & BZQ & 1.5595 & LSP \\
CTA 102 & FSRQ & [HB89] 2230+114 & BZQ & 1.037 & LSP \\
RBS 1895 & bll & RBS 1895 & Beamed AGN & 0.194 & HSP \\
1RXS J225146.9-320614 & bcu & 1RXS J225146.9-320614 & BZB & 0.246 & HSP \\
3C 454.3 & FSRQ & 3C 454.3 & BZQ & 0.859 & LSP \\
PKS 2325+093 & fsrq & PKS 2325+093 & BZQ & 1.843 & LSP \\
PKS 2331-240 & bcu & PKS 2331-240 & BZQ & 0.0477 & LSP \\
1ES 2344+514 & bll & 2MASX J23470479+5142179 & BZB & 0.044 & HSP \\
H 2356-309 & bll & H 2356-309 & BZB & 0.1651 & HSP \\
PKS 0607-549 & bcu & PKS 0607-549 & Beamed AGN & None & LSP \\
PKS 2145+06 & fsrq & PKS 2145+06 & BZQ & 0.99 & LSP \\
4C +06.41 & fsrq & 4C 06.41 & FSRQ & 1.27 & LSP \\
4C +51.37 & fsrq & 1RXS J174036.3+521155 & Beamed AGN & 1.375 & LSP \\
OG 050 & FSRQ & PMN J0532+0732 & FSRQ & 1.254 & LSP \\
4C +73.18 & fsrq & 4C +73.18 & BZQ & 0.3021 & None \\
PKS 0426-380 & bll & PKS 0426-380 & Beamed AGN & 1.11 & LSP \\
S4 1849+67 & FSRQ & 8C 1849+670 & Beamed AGN & 0.657 & LSP \\
NRAO 140 & fsrq & 4C +32.14 & BZQ & 1.258 & LSP \\
PMN J2010-2524 & fsrq & 1RXS J201020.0-252356 & BZQ & None & None \\
TXS 0222+185 & fsrq & RBS 0315 & BZQ & 2.69 & LSP \\
RX J062308.0-643619 & fsrq & 2MASX J06230765-6436211 & BZQ & 0.1289 & LSP \\
PKS 2052-47 & fsrq & PKS 2052-47 & BZQ & 1.489 & LSP \\
PKS 2008-159 & fsrq & PKS 2008-159 & BZQ & 1.18 & LSP \\
B2 0920+39 & bcu & B2 0920+39 & Beamed AGN & None & LSP \\
B2 2023+33 & bll & B2 2023+33 & Beamed AGN & 0.219 & LSP \\
1E 1415.6+2557 & bll & [HB89] 1415+259 & BZB & 0.237 & HSP \\
PKS 0834-20 & fsrq & [HB89] 0834-201 & Beamed AGN & 2.752 & LSP \\
MG4 J162551+4346 & fsrq & 87GB 162418.8+435342 & BZQ & 1.048 & None \\
\end{longtable*}
\tablenotetext{a}{Redshifts used are from \citet{Lien_2025}. If redshifts are not available in the Swift-BAT 157-Month catalog, we use the redshift listed in \citet{Ajello_2022}}

\section{Review of Blazar Luminosity Functions}
\label{sec:lf_review}

In this appendix, we present a review of the formalism for the luminosity-dependent density evolution (LDDE) luminosity function models used to make estimates for the cumulative source counts for future MeV $\gamma$ ray instruments. The parameters for the luminosity functions used in this work, \citet{Rajguru2025GammaRayLF}, \citet{Toda2020FSRQsBAT}, and \citet{Ajello_2014}, are shown in Table \ref{tab:LF_params}. 

\subsection{LDDE Luminosity Function Formalism}

In these luminosity functions, the local LF, where z=0, is described by a smoothed-broken-power-law, shown in Equation \ref{eq:LF_no_gamma}. A is the normalization, $\gamma_1$ and $\gamma_2$ are the indices of the power laws representing the low and high components of the LF, and $L_*$ is the break luminosity.

\begin{equation}
    \Phi(L_{\gamma}, V(z=0)) = \frac{A}{\text{ln}(10)L_{\gamma}} \left[ \left( \frac{L_{\gamma}}{L_{*}}\right)^{\gamma_1} + \left( \frac{L_{\gamma}}{L_{*}}\right)^{\gamma_2}\right]^{-1}
    \label{eq:LF_no_gamma}
\end{equation}

\noindent Many luminosity functions, such as those presented in \citet{Rajguru2025GammaRayLF} and \citet{Ajello_2014} also consider the photon index distribution of the sources. In these cases, the local LF can be rewritten as Equation \ref{eq:LF_with_gamma}.

\begin{equation}
    \Phi(L_{\gamma}, V(z=0), \Gamma) = \Phi(L_{\gamma}, V(z=0)) \times \frac{dN}{d\Gamma}
    \label{eq:LF_with_gamma}
\end{equation}

\noindent In both \citet{Rajguru2025GammaRayLF} and \cite{Ajello_2014}, the distribution of the photon index for the sources is written as a Gaussian distribution, shown in Equation \ref{eq:gamma_dist_fn}, where $\sigma$ is the dispersion of the Gaussian, and $\mu(L_{\gamma})$ is the luminosity-dependent mean of the Gaussian, shown in Equation \ref{eq:mean_guass}.

\begin{deluxetable*}{cccc}
\tablecaption{Parameters of Luminosity Functions for selected works \label{tab:LF_params}}
\tablehead{
\colhead{} & \colhead{\shortstack{\citet{Rajguru2025GammaRayLF}\\LAT FSRQ}} & \colhead{\shortstack{\citet{Toda2020FSRQsBAT}\\BAT FSRQ\tablenotemark{e}}} & \colhead{\shortstack{\citet{Ajello_2014}\\LAT BLL}}
}
\startdata
A & (18878.4 $\pm$ 830.0)\tablenotemark{a} & $-13.02^{+0.25}_{-0.35}$\tablenotemark{b} & $3.39^{+7.44}_{-2.13}\times10^4$\tablenotemark{a} \\
$L_{*}$\tablenotemark{c} & (1.09 $\pm$ 0.52) & $10^{3}$ & $0.28^{+0.43}_{-0.21}$ \\
$\gamma_1$ & 0.29 $\pm$ 0.03 & 5.0 & $0.27^{+0.26}_{-0.46}$ \\
$\gamma_2$ & 1.63 $\pm$ 0.24 & $0.80^{+0.05}_{-0.08}$ & $1.86^{+0.86}_{-0.48}$ \\
$\sigma$ & 0.182 $\pm$ 0.005 & -- & $0.26^{+0.02}_{-0.02}$ \\
$\mu^*$ & 2.42 $\pm$ 0.01 & -- & $2.10^{+0.03}_{-0.03}$ \\
$\beta$ & 0.025 $\pm$ 0.008 & -- & $6.46^{+2.34}_{-2.07} \times10^{-2}$ \\
$p_1$  & 3.5 $\pm$ 1.0 & $3.58^{+0.42}_{-3.08}$ & $4.50^{+0.75}_{-0.61}$ \\
$p_2$  & -9.0 $\pm$ 1.4 & $-7.7^{+1.5}_{-0.3}$ & $-7.37^{+2.95}_{-5.43}$ \\
$z_c^*$  & 2.05 $\pm$ 0.016 & $1.36^{+0.64}_{-1.03}$ & $1.34^{+0.22}_{-0.27}$ \\
$L_c$\tablenotemark{d} & $10^{48}$ & $10^{47.5}$ & $10^{48}$ \\
$\alpha$  & 0.22 $\pm$ 0.01 & $0.42^{+0.18}_{-0.06}$ & $4.53^{+4.98}_{-6.52}\times10^{-2}$  \\
$\tau$ & -- & -- & $4.92^{+1.45}_{-2.12}$
\enddata
\tablenotetext{a}{In units of $10^{-13}$ $\mathrm{Mpc^{-3}erg^{-1}s}.$}
\tablenotetext{b}{$\mathrm{log_{10}A}$ in units of $\mathrm{Mpc^{-3}}.$} 
\tablenotetext{c}{In units of $10^{48}$ $\mathrm{erg}$ $\mathrm{s^{-1}}.$}
\tablenotetext{d}{In units of $\mathrm{erg}$ $\mathrm{s^{-1}}.$}
\tablenotetext{e}{Note that in \citet{Toda2020FSRQsBAT}, $\gamma_1$ and $L_c$ are fixed.}
\end{deluxetable*}

\begin{equation}
    \frac{dN}{d\Gamma} = e^{-\frac{(\Gamma-\mu(L_{\gamma}))^2}{2\sigma^{2}}}
    \label{eq:gamma_dist_fn}
\end{equation}

\begin{equation}
    \mu(L_{\gamma}) = \mu^{*} + \beta \times (\text{log}_{10}(L_{\gamma}) - 46)
    \label{eq:mean_guass}
\end{equation}

\noindent In this work, we adopt the luminosity-density dependent evolution (LDDE) form of evolutionary trend, where the density of sources as a function of redshift is dependent on the luminosity class of object. This type of model has been shown to well represent the evolution of both BL Lac and FSRQ type blazars \citep{Ajello_2012, Ajello_2014}. For the LDDE LF at redshift z, the local LF is multiplied by the evolution factor $e(L_{\gamma},z)$, and written as Equation \ref{eq:LDDE_LF}.

\begin{equation}
    \Phi(L_{\gamma}, V(z), \Gamma) = \Phi(L_{\gamma}, V(z=0), \Gamma) \times e(L_{\gamma},z)
    \label{eq:LDDE_LF}
\end{equation}

\noindent The evolution factor $e(L_{\gamma},z)$ is given as Equation \ref{eq:evo_factor}, and similarly described as a smoothed-broken-power-law, with indices $p_1$ and $p_2$. $z_c(L_{\gamma})$ is the luminosity-dependent redshift where the evolution changes from positive to negative. $z_c^*$ is the redshift at which the density of the population of FSRQs with luminosity $10^{48}$ erg/s peaks, and $\alpha$ is the the power law index relating the redshift peak and its luminosity.

\begin{equation}
    e(L_{\gamma},z) = \left[ \left( \frac{1+z}{1+z_{c}(L_{\gamma})} \right)^{-p_{1}} + \left( \frac{1+z}{1+z_{c}(L_{\gamma})}\right)^{-p_2}\right]^{-1}
    \label{eq:evo_factor}
\end{equation}

\begin{equation}
    z_c(L_{\gamma}) = z_c^* \cdot (L_{\gamma}/L_{c})^{\alpha}
\end{equation}

In \citep{Ajello_2014}, the slope of the evolutionary index $p_1$ is allowed to vary as a function of a parameter $\tau$ to account for different luminosity classes evolving differently. 

\begin{equation}
    p_1({L_{\gamma})} = p_1^* + \tau \cdot (Log_{10}(L_{\gamma})-46)
\end{equation}

\subsection{K-Corrected Luminosity}

To determine the lower bound of the luminosity integral, we utilize the k-corrected luminosity (Equation \ref{eq:k_corr_lum} calculated using the k-correction from the SED templates constructed in Section \ref{sec:template_SEDS}.

\begin{equation}
    L_{\gamma,min}(S,z) = 4 \pi d_L^2K(z)S
    \label{eq:k_corr_lum}
\end{equation}

The k-correction at a specific energy is determined by:

\begin{equation}
    K(E,z) = \frac{{dL(E)}/{dE}}{dL(E(1+z))dE}
\end{equation}

\bibliographystyle{aasjournal}
\bibliography{paper}

\end{document}